%% file: mscov.tex
\documentclass[12pt]{article}
\pdfoutput=1
\usepackage[ruled,vlined,noresetcount, boxed]{algorithm2e}
\usepackage{amsmath,amssymb,amsthm}
\usepackage{graphicx}
\usepackage{latexsym}
\usepackage{textcomp}
\usepackage{longtable}
\usepackage{multirow, booktabs}
\usepackage{natbib}
\usepackage{url}
\usepackage{hyperref}
\usepackage{amsmath}
\usepackage{mathtools}
\usepackage{chemarrow}
\usepackage{subfigure}
\usepackage{amsmath}
\usepackage{mathtools}
\usepackage{rotating}
\usepackage{multirow}
\usepackage{tikz}
\usepackage[margin=.75in]{geometry}

\newcommand{\tr}{\text{tr}}
\newcommand{\etr}{\text{etr}}
\newcommand{\Exp}[1]{{\text{E}}[ \ensuremath{ #1 } ]  }

\newtheorem{theorem}{Theorem}
\newtheorem{conjecture}{Conjecture}

\ifdefined\revision
\newcommand{\edits}[1]{{\textsf{\textcolor{red}{#1}}}}
\else
\newcommand{\edits}[1]{#1}
\fi

\begin{document}

\title{Shared Subspace Models for Multi-Group Covariance Estimation
  \protect\thanks{Alexander M. Franks is an assistant professor in the Department of Statistics and Applied Probability at the University of California, Santa Barbara 
    (\href{mailto:amfranks@uw.edu}{afranks@pstat.ucsb.edu}).  Peter D. Hoff
    is a Professor in the Department of Statistical Science at Duke University
    (\href{mailto:peter.hoff@duke.edu}{peter.hoff@duke.edu}).
This work was partially supported 
 by the Washington Research Foundation Fund for Innovation in
 Data-Intensive Discovery, the Moore/Sloan Data Science Environments
 Project at the University of Washington, NSF grant DMS-1505136, and NIH grant R03-CA211160-01.
 The authors are grateful to Dr.\ Daniel Promislow (Department of
 Pathology, University of Washington), Dr.\ Jessica Hoffman
 (University of Alabama at Birmingham) and Dr. Julie Josse (INRIA)
 for sharing data and ideas which contributed to framing of this
 paper.}}  \author{Alexander Franks and Peter Hoff} \date{\today}
\maketitle 

\begin{abstract}

  We develop a model-based method for evaluating heterogeneity among
  several $p\times p$ covariance matrices in the large $p$, small $n$ setting.
  This is done by assuming a spiked covariance model for each group
  and sharing information about the space spanned by the group-level
  eigenvectors.  We use an empirical Bayes method to identify a low-dimensional
  subspace which explains variation across all groups and use an MCMC
  algorithm to estimate the posterior uncertainty of eigenvectors and
  eigenvalues on this subspace.  The implementation and utility of our
  model is illustrated with analyses of high-dimensional multivariate
  gene expression. 
\vspace{2em}

\vfill
\end{abstract}

\newpage

\section{Introduction}


Multivariate data can often be partitioned into
groups, each of which represent samples from populations with
distinct but possibly related distributions.  Although historically the
primary focus has been on identifying mean-level differences between
populations, there has been a growing need to identify differences in
population covariances as well.  For instance, in case-control studies, mean-level effects may be small relative to subject
variability; distributional differences between groups may still be
evident as differences in the covariances between features.  Even when
mean-level differences are detectable, better estimates of the
covariability of features across groups may lead to an improved
understanding of the mechanisms underlying these apparent mean-level
differences.  Further, accurate covariance estimation is an essential
part of many prediction tasks (e.g. quadratic discriminant analysis).
Thus, evaluating heterogeneity between covariance
matrices can be an important complement to more traditional analyses
for estimating differences in means across
groups.

To address this need, we develop a novel method for
multi-group covariance estimation.  Our method exploits the fact that
in many natural systems, high dimensional data is often very
structured and thus can be best understood on a lower dimensional
subspace. For example, with gene expression data, \edits{we may be interested how the covariability between expression levels differs in subjects with and without a particular disease phenotype (e.g, how does gene expression covariability differ in different subtypes of leukemia? See Section \ref{sec:app})}.  In these applications, the effective dimensionality is thought to scale with the number of gene regulatory
modules, not the number of genes themselves \citep{Heimberg2016}.  As
such, differences in gene expression across groups should be expressed
in terms of differences between these regulatory modules rather than
strict differences between expression levels.  Such differences can be
examined on a subspace that reflects the correlations resulting from
these modules.  In contrast to most existing approaches for group
covariance estimation, our approach is to directly infer such
subspaces from groups of related data.


Some of the earliest approaches for multi-group covariance estimation
focus on estimation in terms of spectral decompositions.
\cite{Flury1987} developed estimation and testing procedures for the
``common principal components'' model, in which a set of covariance
matrices were assumed to share the same eigenvectors.
\citet{Schott1991, Schott1999} considered cases in which only certain
eigenvectors are shared across populations, and \citet{Boik2002}
described an even more general model in which eigenvectors can be
shared between some or all of the groups.  More recently,
\citet{Hoff2009}, noting that eigenvectors are unlikely to be shared
exactly between groups, introduced a hierarchical model for
eigenvector shrinkage based on the matrix Bingham distribution.  There
has also been a significant interest in estimating covariance matrices
using Gaussian graphical models. For Gaussian graphical models, zeros
in the precision matrix correspond to conditional independence relationships
between pairs of features given the remaining features
\citep{Meinshausen2006}.  \citet{Witten2014} extended existing work in
this area to the multi-group setting, by pooling information about the
pattern of zeros across precision matrices.


Another popular method for modeling relationships between
high-dimensional multivariate data is partial least squares regression
(PLS) \citep{Wold2001}. This approach, which is a special case of a bilinear factor
model, involves projecting the data onto a lower dimensional space which
maximizes the similarity of the two groups.  This technique does not
require the data from each group to share the same feature set.  A
common variant for prediction, partial least squares discriminant
analysis (PLS-DA) is especially common in chemometrics and
bioinformatics \citep{Barker2003}.  Although closely related to the
approaches we will consider here, the primarily goal of PLS-based
models is to create regression or discrimination models, not to
explicitly infer covariance matrices from multiple groups of data.
Nevertheless, the basic idea that data can often be well represented
on a low dimensional subspace is an appealing one that we leverage.

\edits{The high-dimensional multi-group covariance estimation problem we explore in this work is also closely related to several important problems in machine learning.  In particular, it can be viewed as an extension of distance metric learning methods \citep{bellet2012similarity, wang2015survey} to the multiple-metric setting.   Multi-group covariance estimation also has applications in multi-task learning \citep{zhang2016multi, liu2009multi}, manifold and kernel learning tasks \citep{kanamori2012non}, computer vision \citep{vemulapalli2013kernel, pham2008robust} and compressed sensing and signal processing \citep{romero2016compressive}. Recently, covariance matrix and subspace learning has been used in deep learning applications \citep{huang2017riemannian}.}

In this paper we propose a multi-group covariance estimation model by
sharing information about the subspace spanned by group-level
eigenvectors.  Our approach is closely related to
the covariance reducing model proposed by \citet{cook2008covariance},
but their model is applicable only when $n \gg p$.
In this work we focus explicitly on high-dimensional inference in the context
of the ``the spiked covariance model'' (also known as the ``partial
isotropy model''), a well studied variant of the factor model
\citep{Mardia1980, Johnstone2001}.  Unlike most previous methods for multi-group covariance estimation, our shared subspace model can be used to improve high-dimensional covariance estimates, facilitates exploration and interpretation of differences between
covariance matrices, and incorporates uncertainty quantification.  It is also straightforward to integrate assumptions used in previous approaches (e.g. eigenvector shrinkage) to the shared subspace model.

In Section \ref{sec:shared} we briefly review
the behavior of spiked covariance models for estimating a single
covariance matrix and then introduce our extension to the multi-group
setting.  In Section \ref{sec:inference} we describe an efficient empirical
Bayes algorithm for inferring the shared subspace and estimating the
posterior distribution of the covariance matrices of the data
projected onto this subspace. In Section \ref{sec:simulation} we investigate the behavior of this
class of models in simulation and demonstrate how the shared subspace
assumption is widely applicable, even when there is little similarity
in the covariance matrices across groups.  In particular, independent
covariance estimation is equivalent to shared subspace estimation with
a sufficiently large shared subspace.  In Section \ref{sec:asymp} we
use an asymptotic approximation to describe how shared subspace
inference reduces bias when both $p$ and $n$ are large.  Finally, In
Section \ref{sec:app} we demonstrate the utility of a shared subspace
model in an analysis of gene expression data from juvenile leukemia
patients .  Despite the large feature size ($p > 3000$) relative to the
sample size ($n < 100$ per group), we identify interpretable
similarities and differences in gene covariances on a low dimensional
subspace. 

\section{A Shared Subspace Spiked Covariance Model}
\label{sec:shared}


Suppose a random matrix $S$ has a possibly degenerate Wishart$(\Sigma,n$)
distribution with density given by
\begin{equation} 
p(S | \Sigma, n) \propto l(\Sigma: S) =  |\Sigma|^{-n/2} \etr( - \Sigma^{-1}  S/2 ) ,  
\label{eqn:lik}
\end{equation}
\noindent where \edits{etr is the exponentiated trace, the covariance matrix is a positive definite matrix,} i.e. $\Sigma \in \mathcal S_p^+$,  and $n$ may be less than $p$.  Such
a likelihood results from $S$ being, for example, a residual sum
of squares matrix from a multivariate regression analysis. In this
case, $n$ is the number of independent observations minus the rank of
the design matrix.

In this paper we consider multi-group
covariance estimation based on $K$ matrices, $Y_1, ..., Y_K$, where
$Y_k$ is assumed to be an $n_k$ by $p$ matrix of mean-zero normal
data, typically with $n_k \ll p$.  Then, $Y_k^TY_k = S_k$ has a
(degenerate) Wishart distribution as in Equation \ref{eqn:lik}.  To
improve estimation, we seek estimators of each covariance matrix,
$\hat{\Sigma}_k$, that may depend on data from all groups.  Specifically, we posit that the
covariance matrix for each group can be written as
\begin{equation}
\Sigma_k = \sigma^2_k(V\Psi_kV^T + I),
\label{eqn:sspsi}
\end{equation}




\begin{figure}[t]
    \centering
    \subfigure[Projection in $\mathbb{R}^3$]{
    \label{fig:3dplot}
    \includegraphics[width=0.25\textwidth]{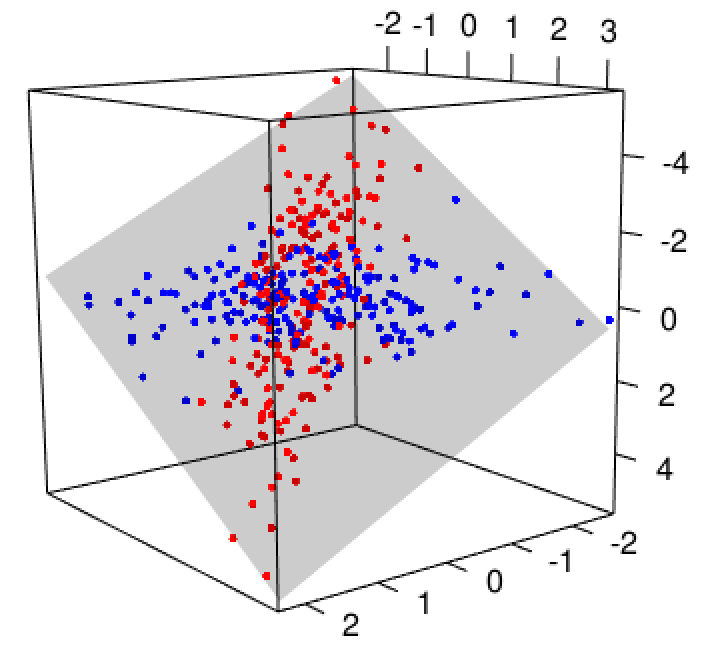}}
\quad
    \subfigure[$Y_kV$]{
    \label{fig:2dscatter}
    \includegraphics[width=0.25\textwidth]{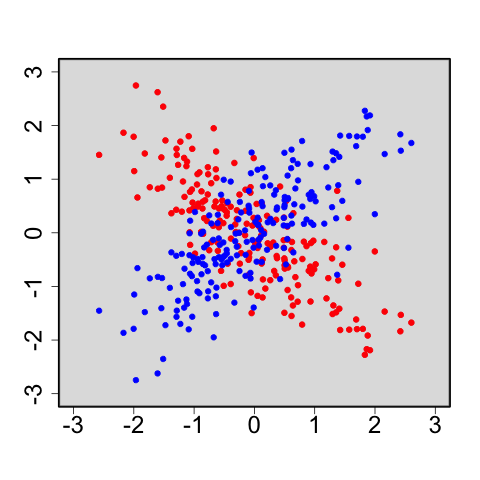}}
\quad 
\subfigure[$Y_kV_{\perp}$]{
\label{fig:2dscatterorth}
\includegraphics[width=0.25\textwidth]{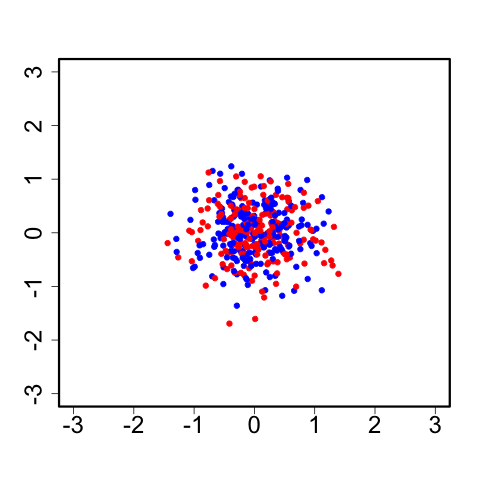}}
\caption{ Two groups of four-dimensional data (red and blue)
      projected into different subspaces.  a) To visualize $Y_k$ we
      can project the data into $\mathbb{R}^3$.  In this illustration, the
      distributional differences between the groups are confined to a
      two-dimensional shared subspace ($VV^T$, grey plane).  b) The
      data projected onto the two-dimensional shared subspace,
      $Y_kV$, have covariances $\Psi_k$ that differ between
      groups. c) The orthogonal projection, $Y_kV_{\perp}$
      has isotropic covariance, $\sigma_k^2I$, for all groups.  }
\label{fig:shared}
\end{figure}

%
\noindent where  $V$ is a $p \times s$ semi-orthogonal matrix whose columns form the basis vectors for subspace of variation shared by all groups. \edits{$\Psi_k$ is a non-isotropic $s \times s$ covariance matrix for each group on this subspace of variation and it is assumed that $s \ll p$.}

\edits{Our model extends the spiked principal components model (spiked PCA)}, studied extensively by \citet{Johnstone2001} and others, to the multi-group setting.  Spiked PCA assumes that
\begin{equation} 
\Sigma = \sigma^2 ( U  \Lambda  U^T  + I )
\label{eqn:spiked}
\end{equation}
\noindent where for $s \ll p$, $\Lambda$ is an $s\times s$ diagonal
matrix and $U \in \mathcal V_{p,s}$, where $\mathcal V_{p,s}$ is the
Stiefel manifold consisting of all $p \times s$ semi-orthogonal matrices in $\mathbb{R}^p$, so that
$U^TU = I_s$.  The spiked covariance formulation is appealing because it
explicitly partitions the covariance matrix into a tractable low rank
``signal'' and isotropic ``noise''.


Classical results for parametric models (e.g., \citet{Schwartz1965})
imply that asymptotically in $n$ for fixed $p$, an estimator will be
consistent for a spiked population covariance as long as the assumed
number of spikes (eigenvalues larger than $\sigma^2$) is greater than
or equal to the true number.  However, when $p$ is large relative to
$n$, as is the case for the examples considered here, things are more
difficult.  Under the spiked covariance model, it has been shown that
if $p/n \rightarrow \alpha >0$ as $n\rightarrow \infty$, the $k$th
largest eigenvalue of $S/(n\sigma^2)$ will converge to an upwardly
biased version of $\lambda_{k}+1$ if $\lambda_k$ is greater than
$\sqrt{\alpha}$ \citep{Baik2006, Paul2007}.  This has led several
authors to suggest estimating $\Sigma$ via shrinkage of the
eigenvalues of the sample covariance matrix. In particular, in the
setting where $\sigma^2$ is known, \citet{Donoho2013} propose
estimating all eigenvalues whose sample estimates are smaller than
$\sigma^2(1+\sqrt{\alpha})^2$ by $\sigma^2$, and shrinking the larger
eigenvalues in a way that depends on the particular loss function
being used.  These shrinkage functions are shown to be asymptotically
optimal in the $p/n\rightarrow \alpha$ setting.

Single-group covariance estimators of the spiked PCA form are equivariant with
respect to rotations and scale changes, but the situation should be
different, when we are interested in estimating multiple
covariance matrices from distinct but related groups with shared features.  Here,
equivariance to distinct rotations in each group is an unreasonable assumption;
both eigenvalue \emph{and} eigenvector shrinkage can play an important
role in improving covariance estimates.  

\edits{In the multi-group setting, we account for similarity between group-level eigenvectors by positing that the anisotropic variability from each group occurs on a
common low dimensional subspace.}  Throughout this
paper we will denote to the shared subspace as
$VV^T \in \mathcal G_{p,s}$, where $\mathcal G_{p,s}$ is the
Grassmannian manifold consisting of all $s$-dimensional linear subspaces
of $\mathbb{R}^p$ \citep{Chikuse2012}.  Although $V$ is only
identifiable up to right rotations, the matrix $VV^T$, which defines
the plane of variation shared by all groups, is identifiable for a
fixed dimension, $s$.  To achieve the most dimension reduction, we target the shared subspace of
minimal dimension, e.g. the shared subspace for which all
$\Psi_k$ are full rank.  Such a minimal subspace is known as the
\emph{central} subspace \citep{cook2009regression}.   Later, to emphasize the connection to the spiked PCA model
(\ref{eqn:spiked}), we will write $\Psi_k$ in terms of its
eigendecomposition, $\Psi_k = O_k\Lambda_kO_k$, where $O_k$ are
eigenvectors and $\Lambda_k$ are the eigenvalues of $\Psi_k$ (see
Section \ref{sec:bayes}).

For the shared subspace model,
$V^T\Sigma_kV = \sigma_k^2(\Psi_k + I)$ is an anisotropic
$s$-dimensional covariance matrix for the projected data, $Y_kV$.  In
contrast, the data projected onto the orthogonal space,
$Y_kV_{\perp}$, is isotropic for all groups.  In Figure
\ref{fig:shared} we provide a simple illustration using simulated
$4$-dimensional data from two groups.  In this example, the
differences in distribution between the groups of data can be
expressed on a two dimensional subspace spanned by the columns of
$V \in \mathcal{V}_{4, 2}$.  Differences in the correlations between
the two groups manifest themselves on this shared subspace, whereas
only the magnitude of the isotropic variability can differ between
groups on the orthogonal space.  Thus, a shared subspace model
can be viewed as a covariance partition model, where one partition
includes the anisotropic variability from all groups and the other
partition is constrained to the isotropic variability from each group.
This isotropic variability is often characterized as measurement
noise.


\section{Empirical Bayes Inference}
\label{sec:inference}

In this section we outline an empirical Bayes approach for estimating
a low-dimensional shared subspace and the covariance matrices of the
data projected onto this space. As we discuss in Section
\ref{sec:simulation}, if the spiked covariance model holds for each
group individually, then the shared subspace assumption also holds,
where the shared subspace is simply the span of the group-specific
eigenvectors, $U_1, ..., U_K$. In practice, we can usually identify a
shared subspace of dimension $s \ll p$ that preserves most of the
variation in the data.  Our primary objective is to identify the
``best'' shared subspace of fixed dimension $s < p$.  Note that this subspace accounts for the across-group similarity, and thus can be viewed as a hyperparameter in a hierarchical model.  Although a fully Bayesian approach may be preferable in the absence of computational limitations, in this paper we propose computationally tractable empirical Bayes inference.  In the empirical Bayes approach, hyperparameters are first estimated via maximum marginal likelihood, often using the expectation-maximization algorithm \citep{lindstrom1988newton}.  In many
settings such an approach yields group-level inferences that are close to that which would be obtained if the correct across-groups model were known (see for example \citealp{efron1973stein}).
In Section \ref{sec:em} we describe the expectation-maximization algorithm for
estimating the maximum marginal likelihood of the shared subspace,
$VV^T$.  This approach is computationally tractable for
high-dimensional data sets.  Given an inferred subspace, we then seek
estimators for the covariance matrices of the data projected onto this
space.  Because seemingly large differences in the point estimates of
covariance matrices across groups may not actually reflect
statistically significant differences, in Section \ref{sec:bayes} we
also describe a Gibbs sampler that can be used to generate estimates
of the projected covariance matrices, $\Psi_k$, and their
associated uncertainty.  Later, in Section \ref{sec:simulation} we
discuss strategies for inferring an appropriate value for $s$ and
explore how shared subspace models can be used for exploratory data
analysis by visualizing covariance heterogeneity on two or three
dimensional subspaces.



\subsection{Estimating the  Shared Subspace}
\label{sec:em}

In this section we describe a maximum marginal likelihood procedure for
estimating the shared subspace, $VV^T$, based on the
expectation-maximization (EM) algorithm.  The full likelihood
for the shared subspace model can be written as

\begin{align}
\nonumber p(S_1, ... S_k | \Sigma_k,n_k) &\propto \prod_{k=1}^K |\Sigma_k|^{-n_k/2}\etr(-\Sigma_k^{-1}S_k/2)  \\
\nonumber &\propto \prod_{k=1}^K  |\Sigma_k|^{-n_k/2}\etr(-(\sigma_k^2(V\Psi_kV^T +
  I))^{-1}S_k/2) \\
\nonumber &\propto \prod_{k=1}^K  |\Sigma_k|^{-n_k/2}\etr(-\left[V(\Psi_k +
  I)^{-1}/\sigma_k^2 V^T + (I-VV^T)/\sigma^2_k\right]S_k/2)
  \\
&\propto \prod_{k=1}^K  (\sigma_k^2)^{-n_k(p-s)/2}|M_k|^{-n_k/2}\etr(-\left[VM_k^{-1}V^T + \frac{1}{\sigma^2_k} (I-VV^T)\right]S_k/2) ,
\end{align}
\noindent where we define $M_k = \sigma_k^2(\Psi_k + I)$.  The log-likelihood in
$V$ (up to an additive constant) is
\begin{align}
\nonumber l(V) &= \sum_k \tr\left(-(VM_k^{-1}V^T +    VV^T/\sigma^2_k)S_k/2\right)\\
&=\frac{1}{2}\sum_k \tr\left((\frac{1}{\sigma_k^2}I-M_k^{-1})V^T
  S_kV\right).
\label{eqn:likV}
\end{align}
We maximize the marginal likelihood of $V$ with an EM algorithm, where
$M_k^{-1}$ and $\frac{1}{\sigma_k^2}$ are considered the
``missing'' parameters.  We assume independent Jeffreys
prior distributions for both $\sigma_k^2$ and $M_k$.  The Jeffreys prior
distributions for these quantities correspond to
$p(\sigma_k^2) \propto 1/\sigma_k^2$ and
$p(M_k) \propto |M_k|^{-(s+1)/2}$.  From the likelihood it can easily
be shown that the conditional posterior for $M_k$ is
$$p(M_k | V) \propto |M_k|^{-(n_k + s + 1)/2}\etr(-(M_k^{-1}V^TS_kV)/2) $$
\noindent which is an inverse-Wishart($V^TS_kV$, $n_k$) distribution.  The
conditional posterior distribution of $\sigma_k^2$ is simply
$$p\left(\sigma^2 | V\right) \propto (\sigma_k^2)^{-n_k(p-s)/2-1}\etr\left(- (I-VV^T)S_k/[2\sigma_k^2]\right)  $$
\noindent which is an inverse-gamma($n_k(p-s)/2$,
$\text{tr}[(I-VV^T)S_k]/2$) distribution.  We summarize our approach in Algorithm \ref{alg:em} below.  



\begin{algorithm}[t]
\SetAlgoLined
 Initialize $V_0 \in \mathcal V_{p,s}$\;
    
 \While{$||V_t - V_{t-1}||_F > \epsilon$}{
    \textbf{E-step:} \\
      \For{$k \gets 1$ \textbf{to} $K$}{
        $\phi^{(k)}_t \gets E[M_k^{-1} | V_{(t-1)}] = n_k(V_{(t-1)}^T S_kV_{(t-1)})^{-1}$\; 
        $\tau^{(k)}_t \gets E[\frac{1}{\sigma_k^2}|V_{(t-1)}] = \frac{n_k(p-s)}{\text{tr}[(I-V_{(t-1)}V_{(t-1)}^T)S_k]}$\;
      }
    \textbf{M-step:}\\
    \quad $V_{t} \gets \underset{V \in \mathcal V_{p,s}}{\text{arg } \text{max }}  \sum_k \text{tr}\left(-(V\phi^{(k)}_tV^T + \tau^{(k)}_tVV^T)S_k/2\right)$\;
 }
 \caption{Shared Subspace EM Algorithm}
 \label{alg:em}
\end{algorithm}

  

For the M-step, we use a numerical algorithm for optimization over the Stiefel manifold.  The algorithm uses the Cayley transform to preserve the orthogonality constraints in $V$ and has computationally complexity that is dominated by the dimension of the shared subspace, not the number of features \citep{Wen2013}. \edits{Specifically, the optimization routine has time complexity O($ps^2 + s^3$), and consequently, our approach is computationally efficient for relatively small values of $s$, even when $p$ is large.  Run times are typically on the order of minutes for values of $p$ as large as 10,000 and moderate values of $s$ (e.g. $< 50$).  See Figure \ref{fig:runtimes} in Appendix B for a plot with typical run times in simulations with a range of values of $p$ and $s$. } 

\paragraph{Initialization and Convergence:}
The Stiefel manifold is compact and the marginal
  likelihood is continuous, so the likelihood is bounded.  Thus, the EM
  algorithm, which increases the likelihood at each iteration, will
  converge to a stationary point \citep{WuEM}. However, maximizing the
  marginal likelihood of the shared subspace model corresponds to a
  non-convex optimization problem over the Grassmannian manifold and
  may converge to a sub-optimal local mode or stationary
  point. Other work involving optimization on the Grassmannian has
  found convergence to non-optimal stationary values problematic and emphasized
  the importance of good (e.g. $\sqrt{n}$-consistent) starting values
  \citep{Cook2016}. Our empirical results on simulated data confirms
  that randomly initialized starting values converge to sub-optimal
  stationary values, and so in practice we initialize the algorithm at
  a carefully chosen starting value \edits{based on the eigenvectors of a pooled covariance estimate.  We give the details for this initialization strategy below.}

First, note that when the shared subspace model holds, the first $s$
  eigenvectors, from \emph{any} of the groups can be used to construct a
  $\sqrt{n}$-consistent estimator of $VV^T$.  In particular, if
  $\hat{U}^{(k)}\hat{U}^{(k)^T}$ is the eigenprojection matrix for the
  subspace spanned by the first $s$ eigenvectors of $S_k$ then it can
  be shown that
  $\sqrt{n} \text{ vec}(\hat{U}^{(k)}\hat{U}^{(k)^T} - VV^T)$
  converges in distribution to a mean-zero normal
  \citep{kollo2000}.  In the large $p$, small $n$ setting, such
  classical asymptotic guarantees give little assurance that the
  resulting estimators would be reasonable, but they nevertheless
  suggest useful strategies for identifying starting value for the EM
  algorithm.

In this work, we choose a subspace initialization strategy
  based on sample eigenvectors of the data pooled from all groups.    Let
  $ Z = \sum_k \pi_k \frac{Z_k}{\sigma_k}$ where $Z_k$ is a mean-zero
  normal with covariance $\Sigma_k$ and $\pi_k = n_k / \sum_k
  n_k$. Then $Z$ is a
  mixture of mean-zero normal distributions with covariance 
  \begin{align*}
\Sigma_Z &= \sum_k \frac{\pi_k}{\sigma^2_k} \Sigma_k\\
  &= V^T( \sum_k \frac{\pi_k}{\sigma_k^2} \Psi_k)V + I,
  \end{align*} Clearly, the first $s$
  eigenvectors of $\Sigma_Z$ span the shared subspace, $VV^T$.  This suggests
  that we can estimate the shared subspace using the scaled and pooled
  data, $Y_{\text{pool}} = [\frac{1}{\sigma_1}Y_1;
  \frac{1}{\sigma_2}Y_2;...;  \frac{1}{\sigma_k}Y_k]$, where
  $Y_{\text{pool}}$ has dimension $(\sum_k n_k) \times p$.  
  We use $\hat{U}_{\text{pool}}\hat{U}_{\text{pool}}^T$  as the
  initial value for subspace estimation algorithm where 
  $\hat{U}_{\text{pool}}$ are the first $s$ eigenvectors of
  $S_{\text{pool}} = Y_{\text{pool}}^TY_{\text{pool}}$.
  If we
  treat $Y_{\text{pool}}$ as an i.i.d. sample from the mixture
  distribution $Z$,  then it is known that  $\hat{U}_{\text{pool}}\hat{U}_{\text{pool}}^T$ is not
  consistent when both $n$ and $p$ growing at the same rate.
For an arbitrary $p$-vector $\eta$, the asymptotic bias of
  $\eta^T\hat{U}_{\text{pool}}\hat{U}_{\text{pool}}^T\eta$ is well
  characterized as a function of the eigenvalues of $\Sigma_Z$ \citep{Mestre2008}.  If either the eigenvalues of $\sum_k
  \frac{\pi_k}{\sigma_k^2} \Psi_k$ or the total sample size $\sum_k n_k$
  are large, $\hat{U}_{\text{pool}}\hat{U}_{\text{pool}}^T$ will
  accurately estimate the shared subspace and likelihood based
  optimization may not be necessary.  However, when either the eigenvalues are
  small or the sample size is small the likelihood based analysis can
  significantly improve inference and $\hat{U}_{\text{pool}}\hat{U}_{\text{pool}}^T$ is a useful starting value for the EM algorithm.



\paragraph{Evaluating Goodness of Fit:}

Tests for evaluating whether eigenvectors from multiple groups
  span a common subspace were explored extensively by
  \citet{Schott1991}.  These tests can be useful for assessing
  whether a shared subspace model is appropriate, but cannot
  be used to test whether a particular subspace explains
  variation across groups.  These results are also based
  on classical asymptotics and are thus less accurate when
  $n \ll p$

Our goodness of fit measure is based on the fact that when $V$ is a
basis for a shared subspace, then for each group, most of the
non-isotropic variation in $Y_k$ should be preserved when projecting
the data onto this space.  To characterize the extent to which this is
true for different groups, we propose a simple estimator for the
proportion of ``signal'' variance that lies on a given subspace.
Specifically, we use the following statistic for the ratio of the sum
of the first $s$ eigenvalues of $V^T \Sigma_k V$ to the sum of the
first $s$ eigenvalues of $\Sigma_k$:
\begin{equation}
\gamma(Y_k: V, \sigma_k^2) = \frac{||Y_kV||^2_F/n_k}{\underset{\widetilde{V} \in \mathcal{V}_{p, s}}{\text{max}}
  ||Y_k\widetilde{V}||^2_F/n_k - B_k }
\label{eqn:ratio}
\end{equation}
\noindent where $||\cdot||_F$ is the Frobenius norm and $B_k$ is a
bias correction where $B_k = \sigma_k^2 p / n_k \sum_k
\left(\frac{m_i^{(k)}}{m_i^{(k)}-\sigma^2_k}\right)$ with $m_i^{(k)}$
the positive solution to the quadratic equation 
\begin{equation}
(m_i^{(k)})^2 + m_i^{(k)}(\sigma^2_kp/n_k - \sigma_k^2 - \hat{\lambda}_i^{(k)}) -
\hat{\lambda}_i^{(k)}\sigma_k^2 = 0.
\label{eqn:quad}
\end{equation}
and $\hat{\lambda}_i^{(k)}$ is the $i$-th eigenvalue of $S_k/n_k$.

\begin{theorem}
Assume $p/n_k \to \alpha_k$ and $s$ is fixed. If
$\Sigma_k = V\Psi_kV^T + \sigma_k^2I$, then $\gamma(Y_k: V, \sigma_k^2)
\overset{a.s}{\to} 1$ as $n_k, p \to \infty$.  
\end{theorem}

\begin{proof}
Since $s$ is fixed and $n_k$ is growing, the numerator,
$||Y_kV||^2_F/n_k$, is a consistent estimator for the sum of the
eigenvalues of $V^T\Sigma_kV$.  In the denominator,
$\underset{\widetilde{V} \in \mathcal{V}_{p,
    s}}{\text{max}}||Y_k\widetilde{V}||^2_F/n_k$
is equivalent to the sum of the first $s$ eigenvalues of the sample
covariance matrix $S_k/n_k$.  \citet{Baik2006} and others have demonstrated that
asymptotically as $p, n_k \rightarrow \infty$ and $p/n_k = \alpha_k$, $\hat{\lambda}^{(k)}_i$ is
positively biased.  Specifically,
\begin{eqnarray}
\hat{\lambda}^{(k)}_i &\overset{a.s.}{\rightarrow}& \lambda^{(k)}_i\left(1 +
                                    \frac{\sigma_k^2\alpha_k}{\lambda^{(k)}_i
                                    - \sigma^2_k}\right) 
\label{eqn:lamAsymp}
\end{eqnarray}
\noindent Replacing $\lambda_i^{(k)}$ by $m_i^{(k)}$ and assuming equality in \ref{eqn:lamAsymp} yields the quadratic equation \ref{eqn:quad}.  The
solution, $m_i^{(k)}$, is an asymptotically (in $n$ and $p$) unbiased
estimator of $\lambda_i^{(k)}$ and
\begin{equation}
\underset{\widetilde{V} \in \mathcal{V}_{p, s}}{\text{max}}
  ||Y_k\widetilde{V}||^2_F/n_k - B_k  \overset{a.s.}{\to} \sum_{i}^s \lambda_i^{(k)} 
\end{equation}
As such, when the shared subspace model holds both the
numerator and denominator of the goodness of fit statistic converge
almost surely to $\sum_{i=1}^{s}
\lambda_i^{(k)}$.  Therefore $\gamma(Y_k: V, \sigma_k^2)
\to 1$.
\end{proof}


The goodness of fit statistic will be close to one for all
groups when $VV^T$ is a shared subspace for the data and typically smaller if
not.  The metric provides a useful indicator of which groups can be
reasonably compared on a given subspace and which groups cannot.  In
practice, we estimate a shared subspace $\hat{V}$ and the isotropic
variances $\hat{\sigma}_k^2$ using EM and compute the plug-in estimate
$\gamma(Y_k: \hat{V}, \hat{\sigma}_k^2)$.  When this statistic is
small for some groups, it may suggest that the rank $s$ of the
inferred subspace needs to be larger to capture the variation in all
groups. If $\gamma(Y_k: \hat{V}, \hat{\sigma}_k^2)$ is substantially
larger than 1 for a particular group, it suggests that the inferred subspace
is too similar to the sample principal components from group $k$.  We
investigate these issues in Section \ref{sec:simulation}, by computing the
goodness of fit statistic for inferred subspaces of different
dimensions on a single data set. In Section \ref{sec:app}, we compute
the estimates for subspaces inferred with real biological data.

\subsection{Inference for Projected Covariance Matrices}
\label{sec:bayes}

The EM algorithm presented in the previous section yields point
estimates for $VV^T$, $\Psi_k$, and $\sigma_k^2$ but does not lead to natural
uncertainty quantification for these estimates.  In this section, we
assume that the subspace $VV^T$ is fixed and known and demonstrate how
we can estimate the posterior distribution for $\Psi_k$.   Note that when
the subspace is known, the posterior distribution of $\Sigma_k$ is
conditionally independent from the other groups, so that we can
independently estimate the conditional posterior distributions for each
group.  


There are many different ways in which we could choose to parameterize
$\Psi_k$.  Building on recent interest in the spiked covariance model
\citep{Donoho2013, Paul2007} we propose a tractable MCMC algorithm by
specifying priors on the eigenvalues and eigenvectors of $\Psi_k$.  By
modeling the eigenstructure, we can now view each covariance
$\Sigma_k$ in terms of the original spiked principal components model.  Equation \ref{eqn:sspsi}, written as a function of
$V$, becomes
\begin{align}
\nonumber \Psi_k &= O_k\Lambda_kO_k^T\\
\Sigma_k &= V\Psi_kV^T + \sigma^2_kI.
\label{eqn:ss}
\end{align}
\noindent Here, we allow $\Psi_k$ to be of rank $r \leq s$ dimensional
covariance matrix on the $s$-dimensional subspace.  Thus, $\Lambda_k$
is an $r \times r$ diagonal matrix of eigenvalues, and
$O_k \in \mathcal{V}_{s,r}$ is the matrix of eigenvectors of
$\Psi_k$.  
For any individual group, this corresponds to the original spiked PCA
model (Equation \ref{eqn:spiked}) with
$U_k = VO_k \in \mathcal{V}_{p, r}$.  Note that the $V$ and $O_k$
  are jointly unidentifiable because for any $s \times s$
  orthonormal matrix $W, VO = VW^TWO = \tilde{V}\tilde{O}$. Once we fix a basis
for the shared subspace, $O_k$ is identifiable.  As such, $O_k$
should only be interpreted relative to the basis $V$, as determined by
the EM algorithm described in Section \ref{sec:em}.  Differentiating the ranks $r$
and $s$ is helpful because it enables us to independently specify a
subspace common to all groups and the possibly lower rank features on
this space that are specific to individual groups.

Although our model
is most useful when the covariance matrices are related across groups,
we can also use this formulation to specify models for multiple
unrelated spiked covariance models.  We explore this in detail in
Section \ref{sec:simulation}.  In Section \ref{sec:app} we introduce a
shared subspace model with additional structure on the eigenvectors
and eigenvalues of $\Psi_k$ to facilitate interpretation of covariance
heterogeneity on a two-dimensional subspace.

The likelihood for $\Sigma_k$ given the sufficient statistic
$S_k = Y_k^TY_k$ is given in Equation \ref{eqn:lik}.  For the
spiked PCA formulation, we must rewrite this likelihood in terms of $V$, $O_k$,
$\Lambda_k$ and $\sigma_k^2$.  First note that by the Woodbury matrix
identity
\begin{align}
\nonumber \Sigma^{-1}_k &=  (\sigma_k^2(U_k\Lambda_kU_k^T+I))^{-1}\\
\nonumber &= \frac{1}{\sigma_k^2}(U_k\Lambda_kU_k^T+I)^{-1}\\
&= \frac{1}{\sigma_k^2}(I-U_k\Omega_kU_k^T),
\end{align}
\noindent where the diagonal matrix $\Omega = \Lambda(I+\Lambda)^{-1}$, e.g. $\omega_i = \frac{\lambda_i}{\lambda_{i}+1}$.  Further, 
\begin{align}
\nonumber |\Sigma_k| &= (\sigma_k^2)^{p}|U_k\Lambda_kU_k^T+I|\\
\nonumber &= (\sigma_k^2)^{p}|\Lambda_k+I| \\
\nonumber &= (\sigma_k^2)^{p}\prod_{i=1}^r(\lambda_i+1)\\
&= (\sigma_k^2)^{p}\prod_{i=1}^r(1-\omega_i),
\end{align}
\noindent where the second line is due to Sylvester's determinant
theorem.  Now, the likelihood of $V$, $O_k$, $\Lambda_k$ and
$\sigma_k^2$ is available from Equation \ref{eqn:lik} by substituting
the appropriate quantities for $\Sigma^{-1}_k$ and $|\Sigma_k|$ and
replacing $U_k$ with $VO_k$:
\begin{equation}
 L(\sigma_k^2, V , O_k \Omega_k : Y_k) \propto
    (\sigma_k^2)^{-n_kp/2}\etr(-\frac{1}{2\sigma_k^2}S_k)\left(\prod_{i=1}^r(1-\omega_{ki})
   \right) ^{n_k/2}
   \etr(\frac{1}{2\sigma_k^2}(VO_k\Omega_kO_k^TV^T)S_k).
\label{eqn:sslik}
\end{equation}
\noindent We use conjugate and semi-conjugate prior distributions for the parameters $O_k$,
$\sigma^2_k$ and $\Omega_k$ to facilitate inference via a Gibbs
sampling algorithm.  In the absence of specific prior information,
invariance considerations suggest the use of priors that lead to
equivariant estimators.  Below we describe our choices for the prior
distributions of each parameter and the resultant conditional posterior
distributions.  \edits{We summarise the Gibbs Sampler in Algorithm \ref{alg:gibbs}}.

\paragraph{Conditional distribution of $\sigma_k^2$:}

From Equation \ref{eqn:sslik} it is clear that the inverse-gamma
class of prior distributions is conjugate for $\sigma_k^2$.  We chose a
default prior distribution for $\sigma^2_k$ that is equivariant with
respect to scale changes.  Specifically, we use the Jeffreys prior distribution, an
improper prior with density $p(\sigma^2_k) \propto 1/\sigma^2_k $.  Under
this prior, straightforward calculations show that the full
conditional distribution of $\sigma_k^2$ is
inverse-gamma$( n_k p/2 , \tr[S_k( I - U_k\Omega_k
U_k^T)/2])$, where $U_k = VO_k$.

\paragraph{Conditional distribution of $O_k$:} Given the likelihood
from Equation \ref{eqn:sslik}, it is easy to show that the class of
Bingham distributions are conjugate for $O_k$ \citep{Hoff2009,
  Hoff2012}.  Again, invariance considerations lead us to use a
rotationally invariant uniform probability measure on
$\mathcal V_{s,p}$.  Under this uniform prior, the full conditional
distribution of $O_k$ has a density proportional to the
likelihood
\begin{align}
\label{lik_vo}
 p(O_k | \sigma^2_k, U_k, \Omega_k) & \propto \etr(\Omega_kO^T_kV^T[S_k/(2\sigma^2_k)]VO_k).
\end{align}
\noindent This is a Bingham$(\Omega, V^T S_k V/(2\sigma^2))$
distribution on $\mathcal V_{s, r}$ \citep{Khatri1977}. A
Gibbs sampler to simulate from this distribution is given in
\citet{Hoff2012}.  

Together, the prior for $\sigma_k^2$ and $O_k$ leads to conditional
(on $V$) Bayes estimators $\hat \Sigma(V^T S_k V)$ that are
equivariant with respect to scale changes and rotations on the
subspace spanned by $V$, so
that $\hat \Sigma(a W V^T S_k V W^T) = a W \hat\Sigma(V^T
S_k V)  W$
for all $a>0$ and $ W\in \mathcal O_{s}$ (assuming an invariant
loss function). Interestingly, if $\Omega_k$ were known (which it is
not), then for a given invariant loss function the Bayes estimator
under this prior minimizes the (frequentist) risk among all
equivariant estimators \citep{Eaton1989}.

\paragraph{Conditional distribution for $\Omega_k$:} Here we specify the
conditional distribution of the diagonal matrix
$\Omega_k = \Lambda_k(I+\Lambda_k)^{-1} = \text{diag}(\omega_{k1},
... \omega_{kr})$.
We consider a uniform(0,1) prior distribution for each element of
$\Omega$, or equivalently, an $F_{2,2}$ prior distribution for the
elements of $\Lambda$.  The full conditional distribution of an
element $\omega_i$ of $\Omega$ is proportional to the likelihood
function
\begin{align}
p(\omega_{ki}|V, O_k, S_k) &\propto_{\omega_{ki}}
  \left(\prod_{i=1}^r(1-\omega_{ki})^{n_k/2}  \right)
  \etr(\frac{1}{2\sigma_k^2}(VO_k\Omega_kO_k^TV^T)S_k) \\
&  \propto  (1-\omega_{ki})^{n_k/2} e^{c_{ki} \omega_{ki}  n_k/2},    
\label{eqn:wpost}
\end{align}
\noindent where $c_{ki} = u_{ki}^T S_k u_{ki}/(n_k \sigma^2_k)$ and $ u_{ki}$ is
column $i$ of $U_k = VO_k$.  \edits{It is straightforward to show that the density for $(1 - \omega_{ki})$ is proportional to a gamma($n_k/2 + 1, c_{ki}n_k/2$) truncated at 1.  Thus, we can easily sample from
this distribution using inversion sampling}.  The behavior of
the distribution for $\omega_{ki}$ is straightforward to understand: if $c_{ki}\leq 1$, then
the function has a maximum at $\omega_{ki} =0$, and decays
monotonically to zero as $\omega_{ki} \rightarrow 1$.  If $c_{ki}>1$ then the
function is uniquely maximized at $(c_{ki}-1)/c_{ki} \in (0,1)$.  To see why
this makes sense, note that the likelihood is maximized when the
columns of $ U_k$ are equal to the eigenvectors of $S_k$
corresponding to its top $r$ eigenvalues
\citep{Tipping1999}. At this value of $U_k$, $c_{ki}$ will then
equal one of the top $r$ eigenvalues of $ S_k/(n_k\sigma_k^2)$.  In the
case that $n_k\gg p$, we expect
$ S_k/(n_k\sigma_k^2)\approx \Sigma_k/\sigma_k^2$, the true (scaled)
population covariance, and so we expect $c_{ki}$ to be near one of the top
$r$ eigenvalues of $\Sigma_k/\sigma^2_k$, say $\lambda_{ki}+1$.  If indeed
$\Sigma_k$ has $r$ spikes, then $\lambda_{ki}>0$,
$c_{ki} \approx \lambda_{ki} +1 > 1$, and so the conditional mode of $w_{ki}$ is
approximately $(c_{ki}-1)/c_{ki} = \lambda_{ki}/(\lambda_{ki}+1)$, the correct value.
On the other hand, if we have assumed the existence of a spike when
there is none, then $\lambda_{ki}=0$, $c_{ki}\approx 1$ and the Bayes estimate
of $w_{ki}$ will be shrunk towards zero, as it should be.  \edits{We  summarise the full Gibbs sampling algorithm below.}

\begin{algorithm}[t]
\SetAlgoLined
 Estimate $\hat V$ using EM (Algorithm \ref{alg:em}).
 Initialize $O_k, \Lambda_k, \sigma_k^2$\;
 \For{$s \gets 1 \textbf{ to number of samples}$}{
     \For{$k \gets 1 \textbf{ to } K$}{
        Sample $\sigma_k^2$ from an inverse-gamma$( n_k p/2 , \tr[S_k( I - \hat V O_k\Omega_k \hat V^T O_k^T)/2])$\;
        Sample $O_k$ from a Bingham($\Omega, \hat V^T S_k \hat V/(2\sigma^2))$\;
        \For{$i \gets 1 \textbf{ to } r$}{
            Sample $(1-\omega_{ki})$ from a gamma($n_k/2 + 1, c_{ki}n_k/2)$ truncated at 1\;
            $\lambda_{ki} \gets \omega_{ki} / (1-\omega_{ki})$
        }   
    }
}
\caption{Gibbs Sampler for Projected Data Covariance Matrices}
\label{alg:gibbs}
\end{algorithm}

\section{Simulation Studies}
\label{sec:simulation}

\begin{figure}[t]
    \centering
    
    \subfigure[Random vs eigen-based initialization]{
      \label{fig:simInitialization}
      \includegraphics[width=0.45\textwidth]{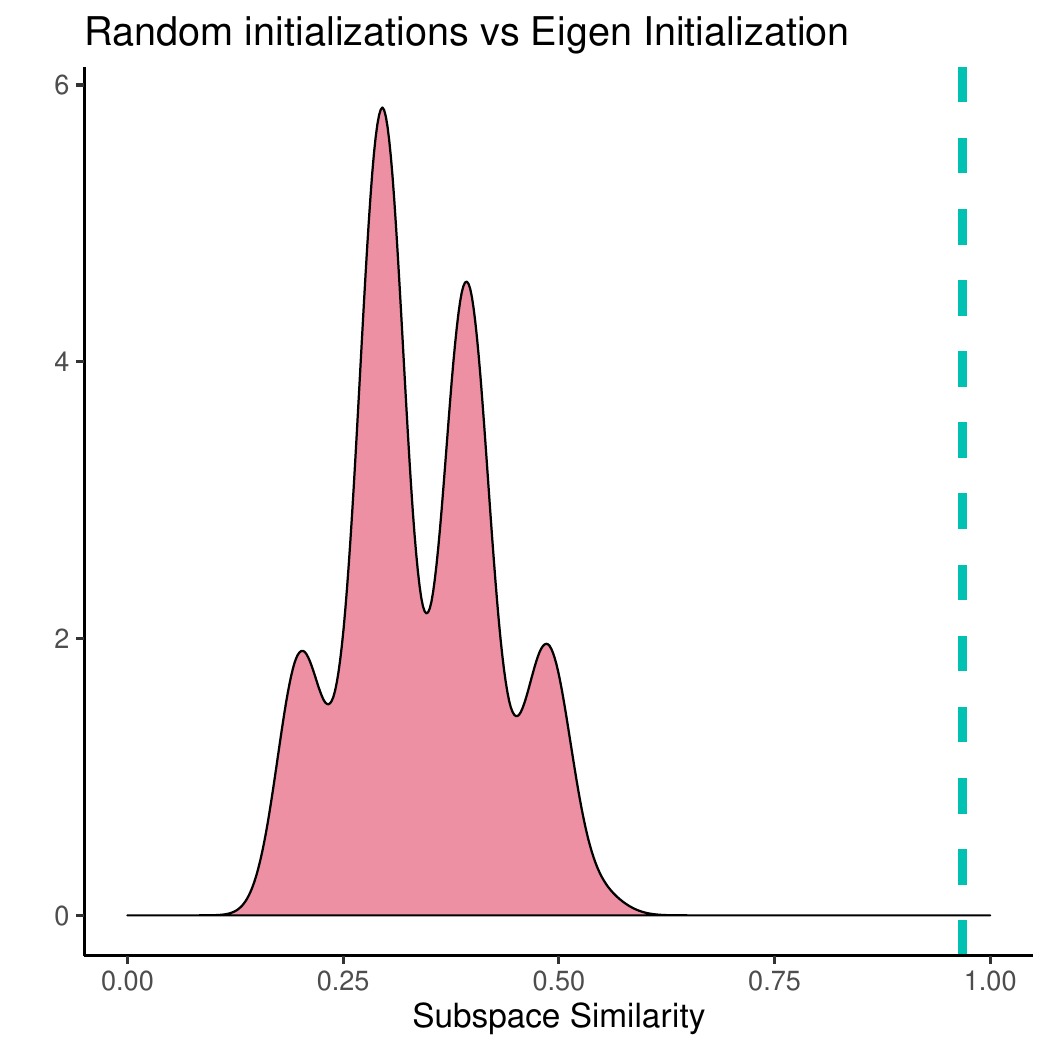}}
    \subfigure[Posterior eigen summaries]{
      \label{fig:simPosterior}
      \includegraphics[width=0.45\textwidth]{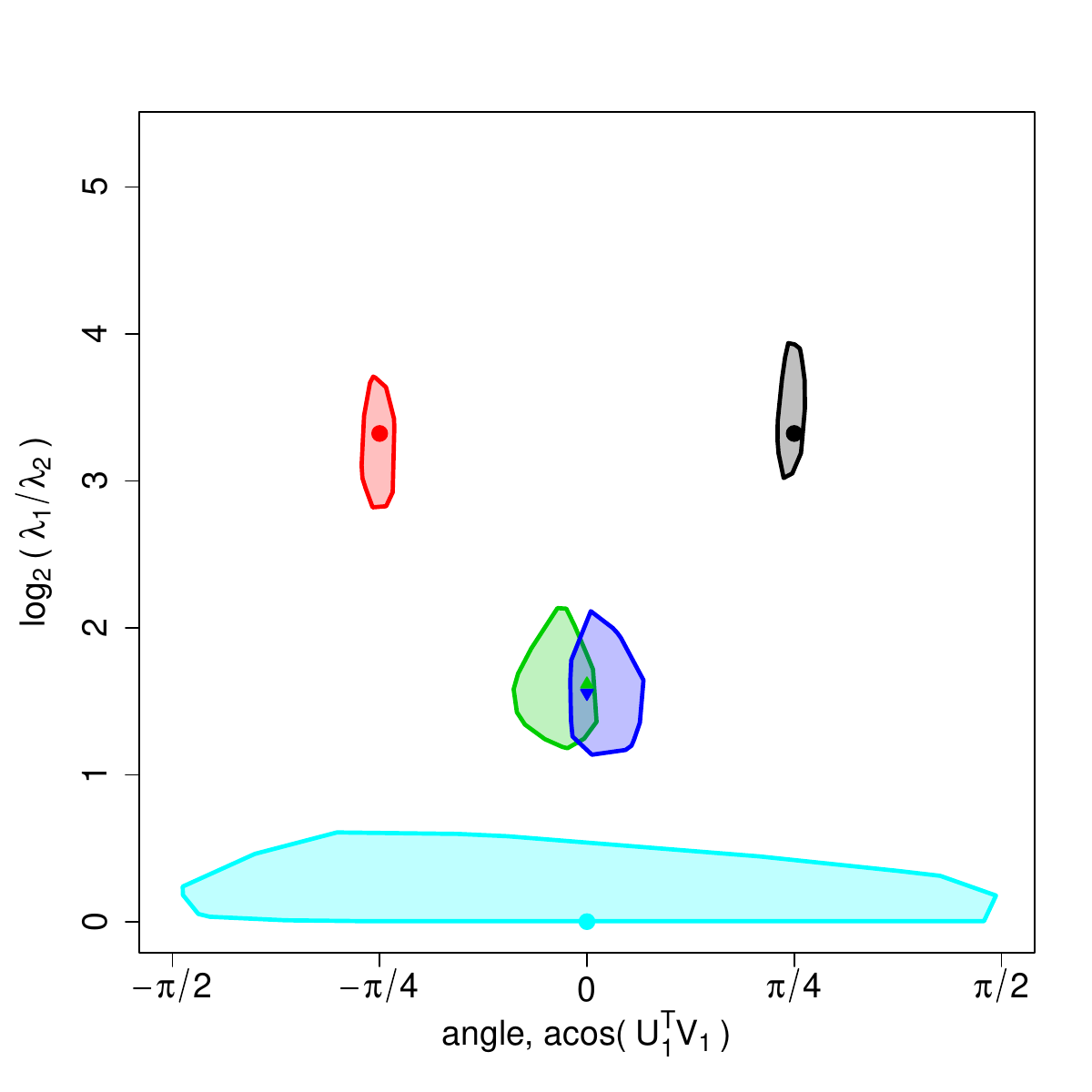}}
    
    \caption{\edits{a) Accuracy of shared subspace estimation, $\tr(\hat V \hat V^T V V^T)/s$ , for randomly initialized (density) and eigen-initialized value of $V$ (dashed line).  If $V$ is initialized uniformly at random from the Stiefel manifold, then typically Algorithm \ref{alg:em} produces a subspace estimate that is sub-optimal.  By contrast, using the initialization strategy described in Section \ref{sec:em}, we achieve excellent accuracy.}   b) 95\% posterior regions for the log of the ratio of
      eigenvalues, $\text{log}(\frac{\lambda_1}{\lambda_2})$, of $\Psi_k$ and the
      orientation of the principal axis on the space spanned by $\hat{V}$
      cover the truth in this simulation.  Dots correspond to true data generating
      parameter values on $\hat{V}^T\Sigma_k\hat{V}$ .  Since $V$
      is only identifiable up to rotation, for this figure we find the Procrustes
      rotation that maximizes the similarity of $\hat{V}$ to the true
      data generating basis. True eigenvalue ratios were 10 (red and
      black), 3 (green and blue) and 1 (cyan).  True
      orientations were $\pi/4$ (black), $-\pi/4$ (red) and
      0 (blue, green, and cyan). \edits{Note that the dark blue and green groups were generated with identical covariance matrices.  Their posterior regions overlap, which suggests that a 95\% region for the difference in eigenvalue ratios and angle would include (0,0)}.  
  \label{fig:sim_figs}
    }

\end{figure}

We start with an example demonstrating how a shared subspace
model can be used to identify statistically significant differences between
covariance matrices on a low dimensional subspace. \edits{Here, we simulate
$K=5$ groups of data from the shared subspace spiked covariance
model with $p=20000$ features, a shared subspace dimension of $s=r=2$, $\sigma_k^2=1$, and $n_k=100$.  We fix the first eigenvalue of $\Psi_k$ from each group to
$\lambda_1=1000$ and vary $\lambda_2$. We generate the basis for the shared subspace and the eigenvectors of $\Psi_k$ by sampling uniformly from the Stiefel manifold. First, in Figure \ref{fig:simInitialization} we demonstrate the importance of the eigen-based initialization strategy proposed in Section \ref{sec:em}. As an accuracy metric, we study the behavior of
$\tr(\hat{V}\hat{V}^TVV^T)/s$ which is bounded by zero and one
and achieves a maximum of one if and only if $\hat{V}\hat{V}^T$
corresponds to the true shared subspace.  In this high dimensional problem, with random initialization, we typically converge to an estimated subspace that has a similarity between 0.25 and 0.5.  With the eigen-based initialization we achieve nearly perfect estimation accuracy ($> 0.95$).}

Next, we summarize estimates of $\Psi_k$ inferred using Algorithm \ref{alg:gibbs} in terms of its eigendecomposition by
computing posterior distributions for the log eigenvalue ratio,
$\text{log}(\frac{\lambda_1}{\lambda_2})$, with $\lambda_1 > \lambda_2$, and the
angle of the first eigenvector on this subspace,
$\text{arctan}(\frac{O_{12}}{O_{11}})$, relative to the first column
of $V$.  In Figure \ref{fig:simPosterior}, we depict the 95\% posterior regions for
these quantities from a single simulation.  Dots correspond to the
true log ratios and orientations of $\hat{V}^T\Sigma_k\hat{V}$, where
$\hat{V}$ is the maximum marginal likelihood for $V$. To compute the
posterior regions, we iteratively remove posterior samples
corresponding to the vertices of the convex hull until only 95\% of
the original samples remain.  Non-overlapping posterior regions
provide evidence that differences in the covariances are
``statistically significant'' between groups.  In this example, the
ratio of the eigenvalues of the true covariance matrices were $10$
(black and red groups), $3$ (green and blue groups) and $1$ (cyan
group).  Larger eigenvalue ratios correspond to more correlated
contours and a value of $1$ implies isotropic covariance.  Note that
for the smaller eigenvalue ratio of $3$, there is more uncertainty
about the orientation of the primary axis.  When the ratio is one, as
is the case for the cyan colored group, there is no information about
the orientation of the primary axis since the contours are spherical.  In this simulation, the 95\% regions all include the true data generating parameters.  As we would hope, we find no evidence of a difference between the blue and green groups, since they have overlapping posterior regions.  This means that a 95\% posterior region for the difference between the groups (0,0), i.e. the model in which the angles and ratios are the same in both groups.  

To demonstrate the overall validity of the shared subspace approach,
we compute the frequentist coverage of these 95\% Bayesian credible
regions for the eigenvalue ratio and primary axis orientation using
one thousand simulations.  For the two groups with eigenvalue ratio
$\lambda_1/\lambda_2 = 3$ the frequentist coverage was close to
nominal at approximately 0.94.  For the groups with
$\lambda_1/\lambda_2 = 10$ the coverage was approximately 0.92.  We
did not evaluate the coverage for the group with
$\lambda_1/\lambda_2 = 1$ (cyan) since this value is on the edge of the
parameter space and is not covered by the 95\% posterior regions
as constructed.  The slight under coverage for the other groups is likely due to the fact
that we infer $VV^T$ using maximum marginal likelihood, and thus
ignore the extra variability due to the uncertainty about the shared
subspace estimate.

\subsection{Rank Selection and Model Misspecification}
\label{sec:misspec}

\begin{figure}[t]
    \centering
    \subfigure[Stein's risk vs $\hat{s}$]{
      \label{fig:sdimension}
      \includegraphics[width=0.3\textwidth]{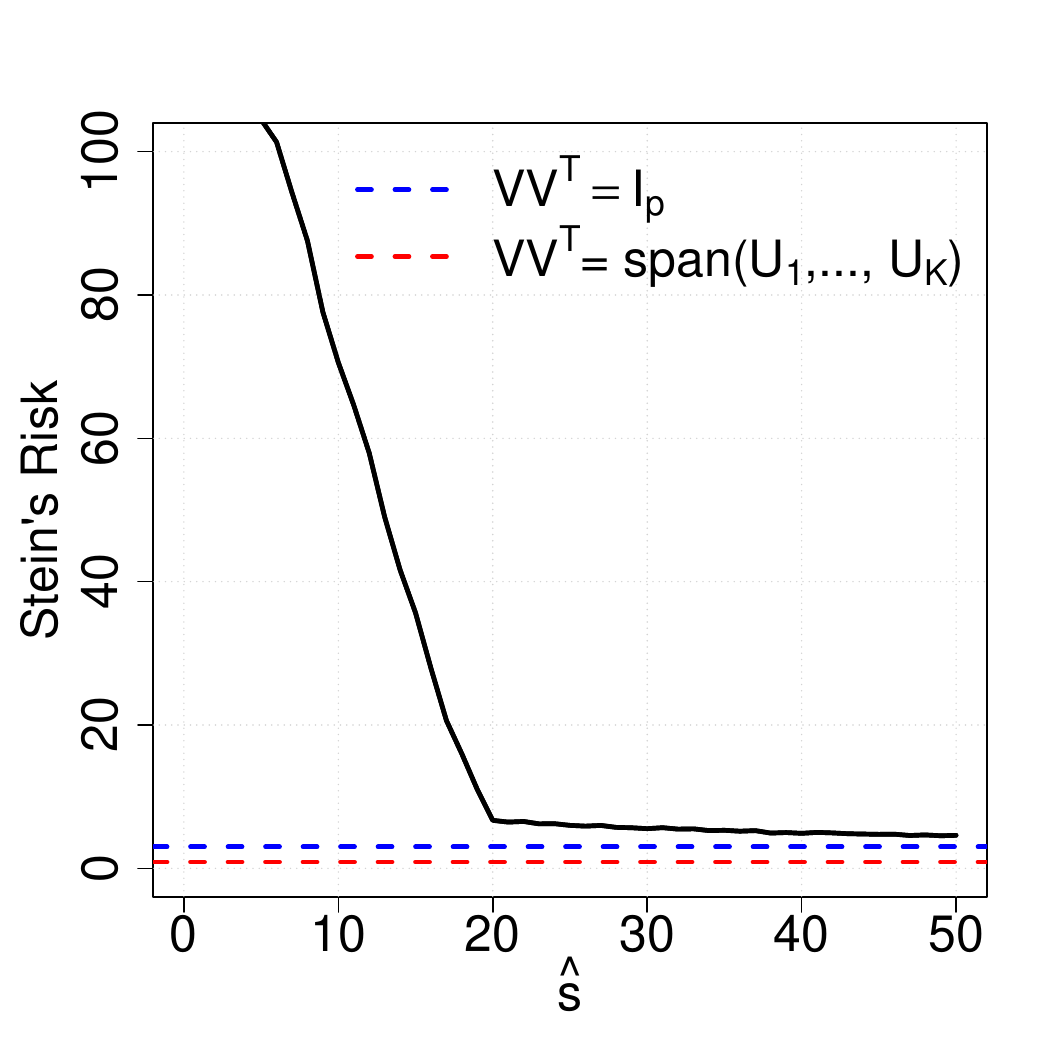}}
    \subfigure[$\hat{s}$ = 5]{
      \label{fig:ratio-s5}
      \includegraphics[width=0.3\textwidth]{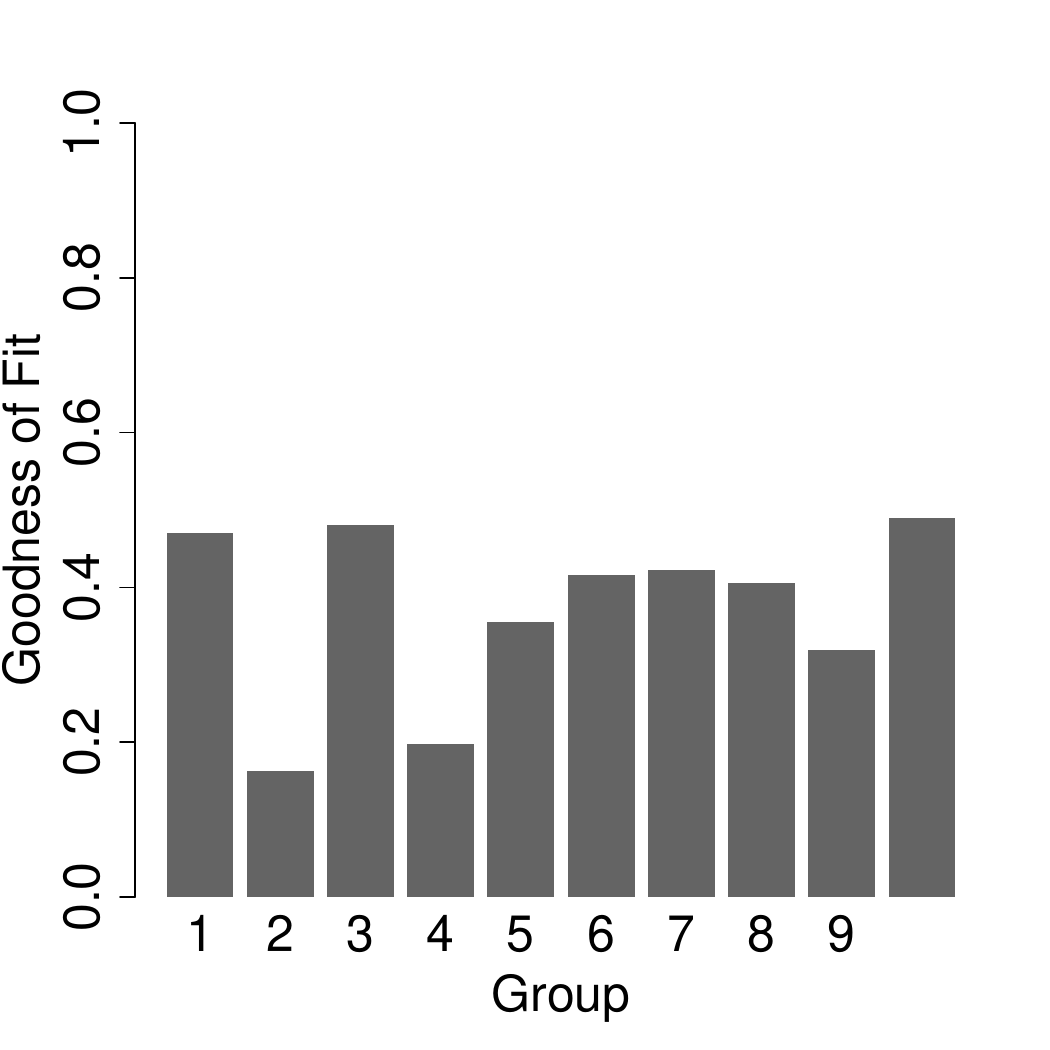}}
    \subfigure[$\hat{s}$ = 20]{
        \label{fig:ratio-s20}
        \includegraphics[width=0.3\textwidth]{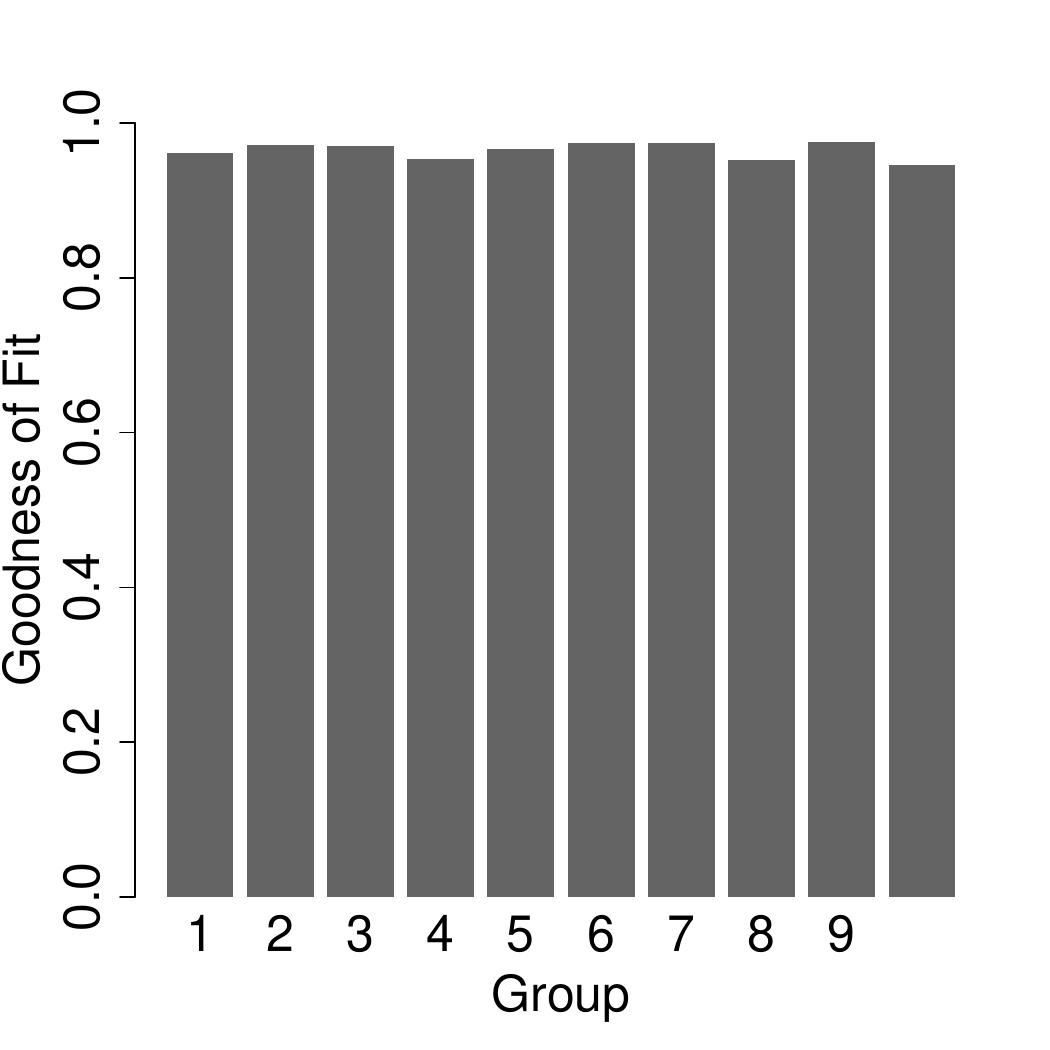}}
      \caption{a) Stein's risk as a function of the
      shared subspace dimension \edits{(solid black line)}.  Data from ten groups, with $U_k$
      generated uniformly on the Stiefel manifold
      $\mathcal{V}_{200, 2}$.  As $\hat{s} \rightarrow p$, the risk converges to the
      risk from independently estimated spiked covariance matrices
      (dashed blue line).  The data also fit a shared subspace model
      with $s=rK$.  If $VV^T = \text{span}(U_1, ..., U_k)$ were known
      exactly, shared subspace estimation yields lower risk than
      independent covariance estimation (dashed red line).  b) For a
      single simulated data set, the goodness of fit statistic,
      $\gamma(Y_k: \hat{V}, \hat{\sigma_k}^2)$, when the assumed
      shared subspace is dimension $\hat{s} = 5$.  c).  For the same
      data set, goodness of fit when the assumed shared subspace is
      dimension $\hat{s} = 20$.  We can capture nearly all of the
      variability in each of the 10 groups using an $\hat{s}=rK=20$
      dimensional shared subspace. }
\label{fig:dimensionPlots}
\end{figure}

Naturally, shared subspace inference works well when the model is
correctly specified.  What happens when the model is not well
specified?  We explore this question \textit{in silico} by simulating data from
different data generating models and evaluating the efficiency of
various covariance estimators.  In all of the following simulations we
evaluate covariance estimates using Stein's loss,
$L_S( \Sigma_k , \hat\Sigma_k) = \text{tr}( \Sigma_k^{-1} \hat
\Sigma_k ) - \log |\Sigma_k^{-1} \Sigma_k | - p$.
Since we compute multi-group estimates, we report the
average Stein's loss
$L(\Sigma_1, ..., \Sigma_K; \hat\Sigma_1, ..., \hat\Sigma_K ) =
\frac{1}{K} \sum_k L_S( \Sigma_k , \hat\Sigma_k)$.
Under Stein's loss, the Bayes estimator is the inverse of
the posterior mean of the precision matrix,
$\hat \Sigma_{k} = \Exp{ \Sigma_k^{-1} | S_k}^{-1}$ which we
estimate using MCMC samples.

We start by investigating the behavior of our model when we
underestimate the true dimension of the shared subspace.  In this
simulation, we generate $K=10$ groups of mean-zero normally
distributed data with $p=200$, $r=2$, $s=p$ and $\sigma_k^2=1$.  We
fix the eigenvalues of $\Psi_k$ to
($\lambda_1, \lambda_2) = (250, 25)$.  Although the signal variance
from each group individually is preserved on a two dimensional
subspace, these subspaces are not similar across groups since the
eigenvectors from each group are generated uniformly from the Stiefel
manifold, $U_k \in \mathcal{V}_{p, r}$.

We use these data to evaluate how well the shared subspace estimator
performs when we fit the data using a shared subspace model of
dimension $\hat{s} < s$.  In Figure \ref{fig:sdimension} we plot
Stein's risk as a function of $\hat{s}$, estimating the risk
empirically using ten independent simulations per value of $\hat{s}$.
The dashed blue line corresponds to Stein's risk for covariance
matrices estimated independently.  Independent covariance estimation
is equivalent to shared subspace inference with $\hat{s} = p$ because
this implies $VV^T = I_p$.  Although the risk is large for small
values of $\hat{s}$, as the shared subspace dimension increases to the
dimension of the feature space, that is $\hat{s} \rightarrow p$, the
risk for the shared subspace estimator quickly decreases.
Importantly, it is always true that rank($[U_1, ..., U_K]) \leq rK$ so
it can equivalently be assumed that the data were generated from a
shared subspace model with dimension $s = rK < p$.  As such, even when
there is little similarity between the eigenvectors from each group,
the shared subspace estimator with $\hat{s} = rK$ will perform well,
provided that we can identify a subspace, $\hat{V}\hat{V}^T$ that is
close to $\text{span}([U_1, ..., U_K])$. When
$\hat{V}\hat{V}^T = \text{span}([U_1, ..., U_K])$ exactly, shared
subspace estimation outperforms independent covariance estimation
(\ref{fig:sdimension}, dashed red line).

From this simulation, it is clear that correctly specifying the
dimension of the shared subspace is important for efficient covariance
estimation.  When the dimension of the shared subspace is too small,
we accrue higher risk.  The goodness of fit statistic,
$\gamma(Y_k: \hat{V}, \hat{\sigma_k}^2)$, can be used to identify when
a larger shared subspace is warranted.  When $\hat{s}$ is too small,
$\gamma(Y_k: \hat{V}, \hat{\sigma_k}^2)$ will be substantially smaller
than one for at least some of the groups, regardless of $\hat{V}$
(e.g. Figure \ref{fig:ratio-s5}).  When $\hat{s}$ is large enough, we
are able to use maximum marginal likelihood to identify a shared
subspace which preserves most of the variation in the data for all
groups (Figure \ref{fig:ratio-s20}).  Thus, for any estimated
subspace, the goodness of fit statistic can be used to identify the
groups that can be fairly compared on this subspace and whether we
would benefit from fitting a model with a larger value of $\hat{s}$.

\edits{Finally, in the appendix, we include a some additional misspecification results.  In particular, we consider two cases in a 10 group analysis: one case in which 7 groups share a common subspace but the other three do not, and a second case in which five groups share one common two dimensional subspace, and the other five groups share a different two dimensional subspace (see Figures \ref{fig:five_share} and \ref{fig:seven_share}).  Briefly, these results indicate that when only some of the groups share a common subspace, we can still usually identify both the existence of the subspace(s) shared by those groups.  We can also identify which groups do not share the space, using the goodness of fit metric.  When there are multiple relevant shared subspaces, we can often identify those distinct modes using a different subspace initialization for the EM algorithm.  }

\paragraph{Model Comparison and Rank Estimation:}

Clearly, correct specification for the rank of the shared subspace is
important for efficient inference.  So far in this section, we have
assumed that the group rank, $r$, and shared subspace dimension, $s$,
are fixed and known.  In practice this is not the case.  Prior to
fitting a model we should estimate these quantities.  Standard
  model selection methods can be applied to select the both $s$
  and $r$.  Common approaches include cross validation and information
  criteria like AIC and BIC.  However, these approaches are
  computationally intensive since they require fitting the model for
  each value of $s$ and $r$.  Here, we estimate the model dimensions
by applying an asymptotically optimal (in mean squared error) singular
value threshold for low rank matrix recovery with noisy data
\citep{Gavish2014}.  This rank estimator is a function of the median
singular value of the data matrix and the ratio $\alpha_k = p/n_k$.
Note that under the shared subspace model, the scaled and
  pooled data described in section \ref{sec:em} can be expressed as
  $Y_{\text{pooled}} = X + Z$ where $V$ are the left singular values
  of $X$ and $Z$ is a noise matrix with zero mean and variance one.
  This is the setting in which \citet{Gavish2014} develop a rank
  estimation algorithm, and so it can be appropriately applied
  to $Y_{\text{pooled}}$ to estimate $s$.

Using this rank estimation approach, we conduct a simulation which
demonstrates the relative performance of shared subspace group
covariance estimation under different data generating models.  We consider three
different shared subspace data models: 1) a low dimensional shared
subspace model with $s=r$; 2) a model in which the spiked covariance
matrices from all groups are identical, e.g.
$\Sigma_k = \Sigma = U\Lambda U^T + \sigma^2I$; and 3) a full rank
shared subspace model with $s=p$.  

We estimate group-level covariance matrices from simulated data using three different variants of the shared
subspace model.  For each of these fits we estimate $r$.  First, we
estimate a single spiked covariance matrix from the pooled data and
let $\hat{\Sigma}_k = \hat{\Sigma}$.  Second, we fit the full rank
shared subspace model.  This corresponds to a procedure in which we
estimate each spiked covariance matrix independently, since $s=p$
implies $VV^T = I_p$.  Finally, we use an ``adaptive'' shared subspace
estimator, in which we estimate both $s$, $r$ and $VV^T$.  

Since full rank estimators do not scale well, we compare the performance of various estimators on a simulated data set with only $p=200$ features.  We also assume for $r=2$ spikes, $\sigma^2_k=1$, and $n_k = 50$.  We fix the non-zero eigenvalues of $\Psi_k$ to $(\lambda_1, \lambda_2) = (250, 25)$.  We simulate 100 independent data sets for each data generating mechanisms.  In Table
\ref{table:groupLoss} we report the average Stein's risk and
corresponding 95\% loss intervals for the estimates derived from each
of these inferential models.

\begin{table}
  \caption[abcd]{Stein's risk (and 95\% loss intervals)
    for different inferential models and data generating models with
    varying degrees of between-group covariance similarity.  For each of $K=10$ groups, we simulate data from three different types of shared subspace models.  For each of these models, $p=200$, $r=2$,
    $\sigma_k^2=1$ and $n_k=50$.  We also fit the data using three
    different shared subspace models: a model in which $s$, $r$ and
    $VV^T$ are all estimated from the data (``adaptive''), a spiked
    covariance model in which the covariance matrices from each group
    are assumed to be identical ($\hat{\Sigma}_k=\hat{\Sigma}$) and a
    model in which we assume the data do \emph{not} share a lower
    dimensional subspace across groups (i.e. $\hat{s} = p$). The estimators
    which most closely match the data generating model have the lowest
    risk (diagonal) but the adaptive estimator performs well relative
    to the alternative misspecified model. \label{table:groupLoss}}
\centering
 \begin{tabular}{ l  l | c | c | c |}
    \multicolumn{2}{c}{} & \multicolumn{3}{c}{\textbf{Inferential Model}} \\
  \multicolumn{2}{c|}{}  & Adaptive & $\hat{\Sigma}_k=\hat{\Sigma}$
                                                           & $\hat{s} = p$ \\  \cline{2-5}
    \multirow{3}{*}{\rotatebox[origin=c]{90}{\textbf{Data Model}}} 
& $s=r=2$ & 0.8 (0.7, 0.9) & 2.1 (1.7, 2.6) & 3.0 (2.9, 3.2) \\ 
   &   $s=r=2$, $\Sigma_k = \Sigma$ & 0.8 (0.7, 0.9) & 0.7 (0.6, 0.8) & 3.0 (2.9, 3.2)\\ 
   &  $s=p=200$ & 7.1 (6.2, 8.0) & 138.2 (119, 153) & 3.0 (2.9, 3.2) \\ 
  \end{tabular}
\end{table}

As expected, the estimates with the lowest risk are derived from the
inferential model that most closely match the data generating
specifications. However, the adaptive estimator has small risk under
model misspecification relative to the alternatives.  For example,
when $\Sigma_k = \Sigma$, the adaptive shared subspace estimator has
almost four times smaller risk than the full rank estimator, in which
each covariance matrix is estimated independently.  When the
data come from a model in which $s=p$, that is, the eigenvectors of $\Psi_k$
are generated uniformly from $\mathcal{V}_{p,r}$, the adaptive
estimator is over an order of magnitude better than the estimator
which assumes no differences between groups.  These results suggest
that empirical Bayes inference for $VV^T$ combined with the rank
estimation procedure suggested by \citet{Gavish2014} can be widely
applied to group covariance estimation because the estimator adapts to
the amount of similarity across groups.  Thus, shared subspace estimation
can be an appropriate and computationally efficient choice when the similarity between
groups is not known \textit{a priori}.  

\edits{Finally, in addition to potential statistical efficiency gains, the empirical Bayes shared subspace estimator has significant computational advantages.  In particular, the total run time for empirical Bayes inference of the shared subspace is significantly smaller than full Bayesian inference for a $p \times r$ dimensional subspace (e.g. Bayesian probabilistic PCA with $s=p$), in particular for larger values of $p$.  Given the difficulty of Bayesian inference on the Stiefel manifold, for large $p$, probabilistic principal component analysis quickly becomes infeasible.  Empirical Bayes inference enables efficient optimization for $\hat V$ and Bayesian inference on the lower dimensional shared subspace (See Figure \ref{fig:runtimes}, Appendix B, for typical run times).} 




\section{Reduction of Asymptotic Bias Via Pooling}
\label{sec:asymp}
Recently, there has been an interest in the asymptotic behavior of
PCA-based covariance estimators in the setting in which
$p, n \to \infty$ with $p/n=\alpha$ fixed.  Specifically, in the
spiked covariance model it is known that when $p$ and $n$ are both
large, the leading eigenvalues of the sample covariance matrix are
positively biased and the empirical eigenvectors form a non-zero angle
with the true eigenvectors \citep{Baik2006, Paul2007}.  Although this
fact also implies that the shared subspace estimators are biased, a
major advantage of shared subspace inference over independent
estimation of multiple covariance matrices is that we reduce the
asymptotic bias, relative to independently estimated covariance matrices, by pooling
information across groups.  The bias reduction appears to be
especially large when there is significant heterogeneity in the first
$s$ eigenvectors of the projected covariance matrices.

Throughout this section we assume $K$ groups of data each with $n_k = n$
observations per group and $s$ a fixed constant.  First, note that if
$\hat{V}\hat{V}^T$ corresponds to the true shared subspace, then
estimates $\hat{\psi}_k$ derived using the methods presented in
Section \ref{sec:bayes} will consistently estimate $\psi_k$ as
$n \to \infty$ regardless of whether $p$ increases as well because
$Y_kV$ has a fixed number of columns.  For this reason, we focus
explicitly on the accuracy of $\hat{V}\hat{V}^T$ (derived using the
maximum marginal likelihood algorithm presented in Section
\ref{sec:em}) as a function of the number of groups $K$ when both $p$
and $n$ are of the same order of magnitude and much larger than $s$.
As an accuracy metric, we again study the behavior of
$\tr(\hat{V}\hat{V}^TVV^T)/s$ which is bounded by zero and one
and achieves a maximum of one if and only if $\hat{V}\hat{V}^T$
corresponds to the true shared subspace.

\begin{conjecture}
Assume that the first $s$ eigenvalues from each of $K$ groups are
  identical with $\lambda_i > \sigma^2(1 + \sqrt{\alpha})$.  Then,
 for $p/n \to \alpha$ and $p, n
  \to \infty$,  $\tr(\hat{V}\hat{V}^TVV^T)/s \overset{a.s.}{\to} \xi$ with  
\begin{equation}1 > \xi \geq  \frac{1}{s}\sum_{i=1}^s  \left(1-\frac{\alpha}{K(\lambda_i - 1)^2}\right) /\left(1 +
    \frac{\alpha}{K(\lambda_i - 1)}\right).
\label{eqn:asympBound}
\end{equation} 
\end{conjecture}

  We prove that the lower bound in \ref{eqn:asympBound} is
in fact achieved when $Y_k$ are identically distributed and show in simulation
that the subspace accuracy exceeds this bound when there is variation
in the eigenvectors across groups. 
In the case of i.i.d. groups, let the covariance matrix $\Sigma_k=\Sigma$ have the shared-subspace form given in
Equation \ref{eqn:sspsi} and without loss of
generality let $\psi_k = \psi$ be a diagonal matrix (e.g assume
the columns of $V$ align with the eigenvectors of $\Sigma$).  In this
case, the complete data likelihood of $V$ (Equation
\ref{eqn:likV}) can be rewritten as
\begin{align*}
\ell(V) &=\frac{1}{2}\sum_k \tr\left((\frac{1}{\sigma^2}I-M^{-1})V^T
  S_kV\right)\\
&=\frac{1}{2}\tr\left(DV^T(\sum_k S_k)V\right).
\end{align*}

\noindent where $\sum_{k=1}^K S_k \sim \text{Wish}(\Sigma, Kn)$.
Since $\psi$ is diagonal and $\sigma^2=1$, 
$M = \sigma^2(\psi + I)$ is diagonal and thus
$D = (\frac{1}{\sigma^2}I-M^{-1})$ is also diagonal with entries
$0 < d_i < 1$ of decreasing magnitude.  

Then, the solution to
$$\hat{V}^{(k)} = \underset{\widetilde{V} \in \mathcal{V}_{p,
    s}}{\text{argmax }} \tr\left(D\widetilde{V}^T
 \sum_k( S_k)\widetilde{V}\right).$$
has $\hat{V}^{(k)}$ equal to the first $s$ eigenvectors of $\sum_k S_k$.  This is
maximized when the columns of $V$ match the first empirical eigenvectors of
$\sum_k S_k$ and has a maximum of $\sum_{i=1}^r d_i\ell_i$ where $\ell_i$ is the
$i$th eigenvalue of $\sum_k S_k$.  Using a result from
\citet{Paul2007}, it can be shown that as long as
$\lambda_i > \sigma^2(1 + \sqrt{\alpha})$ where $\lambda_i$ is the $i$th
eigenvalue of $\Sigma_k$, the asymptotic inner product between
the $i$th sample eigenvector and the $i$th population eigenvector
approaches a limit that is almost surely less than one
$$|\langle\hat{V}_i, V_i\rangle| \overset{a.s.}{\to} \sqrt{\left(1-\frac{\alpha}{K(\lambda_i - 1)^2}\right) /\left(1 +
    \frac{\alpha}{K(\lambda_i - 1)}\right)} $$
%

\noindent As such, we can express asymptotic shared subspace
accuracy for the identical groups model as
\begin{align}
\nonumber \tr(\hat{V}\hat{V}^{^T}VV^T)/s &= \frac{1}{s}\sum_{i=1}^s |\langle\hat{V}_i, V_i\rangle|^2\\
&\overset{a.s.}{\to} \frac{1}{s}\sum_{i=1}^s  \left(1-\frac{\alpha}{K(\lambda_i - 1)^2}\right) /\left(1 +
    \frac{\alpha}{K(\lambda_i - 1)}\right).
\label{eqn:lbound}
\end{align}



Here, the accuracy of the estimate depends on $\alpha$, $K$ and the
magnitude of the eigenvalues, with the bias naturally decreasing as the number
of groups increases.  Most importantly, Equation
\ref{eqn:lbound} provides a useful benchmark for understanding the
bias of shared subspace estimates in the general setting in which
$\psi_k$ varies across groups.  Our conjecture that the subspace
accuracy is larger than the lower bound when the eigenvectors between
groups are variable is consistent with our simulation results.


\begin{figure}[t]
    \centering
    \includegraphics[width=0.7\textwidth]{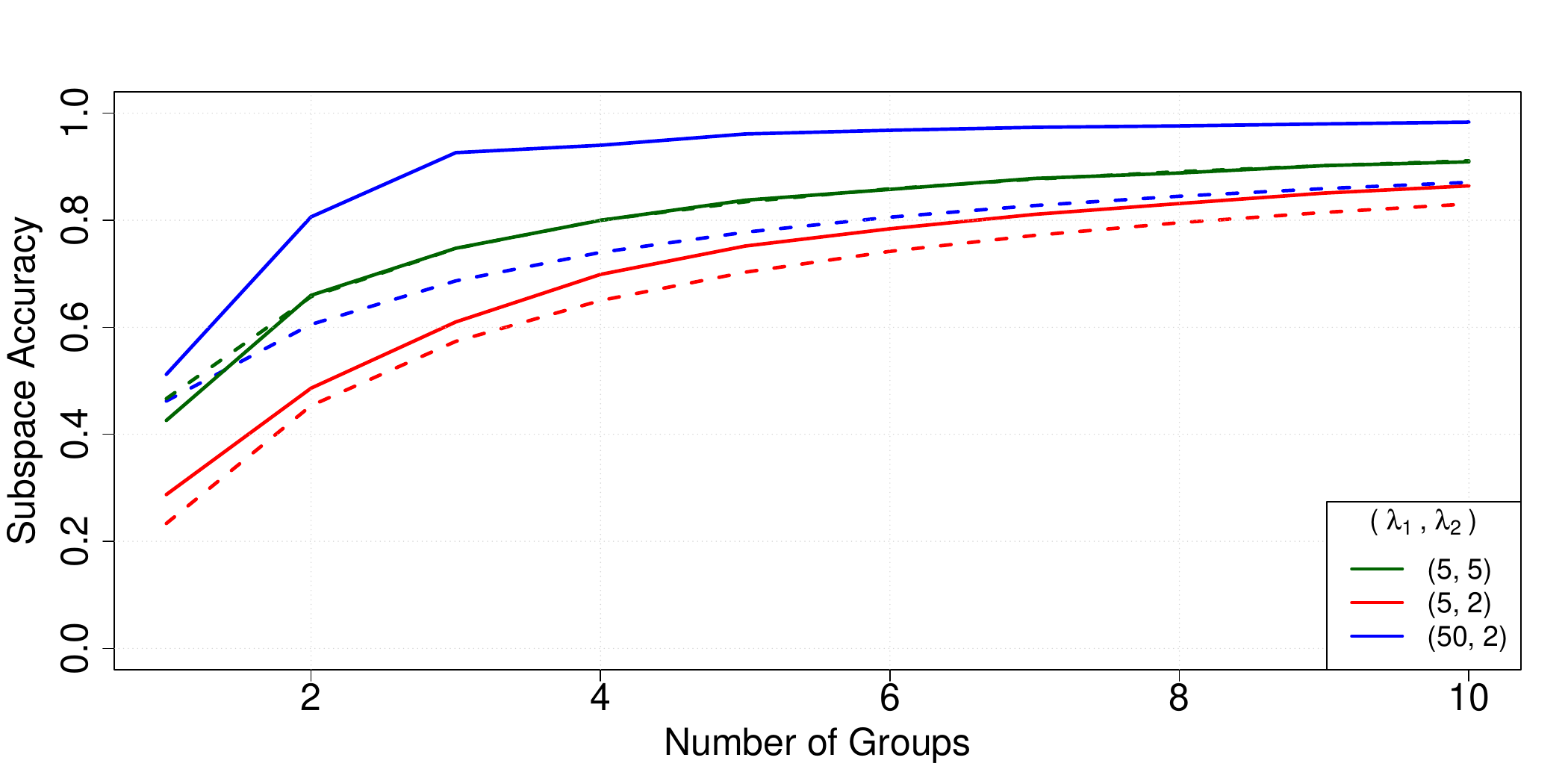}
    \caption{Subspace accuracy $\tr(\hat{V}\hat{V}^TVV^T)/s$ (solid)
      and the asymptotics-based benchmark (dashed)
      as a function of $K$.  When
      $\lambda_1=\lambda_2$ (green), the assumptions used to derive
      the benchmark (identically distributed groups) are met and thus the
      subspace accuracy matches the benchmark.  However, when the
      ratio $\lambda_1/\lambda_2$ is large, the subspace accuracy
      metric can far exceed this benchmark if there is significant
      variation in the eigenvectors of $\psi_k$ across groups.  Small
      increases in accuracy over the benchmark are seen for moderately
      anisotropic data (red) and large increases for highly
      anisotropic data (blue).  }
      \label{fig:asymptotics}
\end{figure}

In Figure \ref{fig:asymptotics} we depict the subspace accuracy metric
$\tr(\hat{V}\hat{V}^TVV^T)/s$ and benchmark
$$\frac{1}{s}\sum_{i=1}^s \left(1-\frac{\alpha}{K(\lambda_i -
    1)^2}\right) /\left(1 + \frac{\alpha}{K(\lambda_i - 1)}\right)$$
for simulated multi-group data generated under the shared subspace
model with $s=2$, $n=50$, $p=200$ and three different sets of eigenvalues.  For each
covariance matrix, the eigenvectors of $\psi_k$ were sampled uniformly
from Stiefel manifold $\mathcal{V}_{2,2}$.  When $\psi_k$ is isotropic
(green) the subspace similarity metric closely matches the benchmark
since the assumptions used to derive this asymptotic result are met.
However, when the eigenvectors of $\psi_k$ vary significantly across
groups and $\lambda_1 \gg \lambda_2$, the subspace accuracy can far
exceed this benchmark (blue).  Intuitively, when the first
eigenvectors of two different groups are nearly orthogonal, each group
provides a lot of information about orthogonal directions on $VV^T$
and so the gains in accuracy exceed those that you would get by
estimating the subspace from a single group with $K$ times the sample
size.  In general the accuracy of shared subspace estimates depends on
the variation in the eigenvectors of $\psi_k$ across groups as well as
the magnitude of the eigenvalues and matrix dimensions $p$ and $n_k$.
Although the shared subspace estimator improves on the accuracy of
individually estimated covariance matrices, estimates can still be
biased when $\alpha$ is very large or the eigenvalues of $\Sigma_k$
are very small for all $k$.  In practice, one should estimate the
approximate magnitude of the bias using the inferred eigenvalues of $\Sigma_k$.  When
these inferred eigenvalues are significantly larger than
$\hat{\sigma}_k^2(1+\sqrt{\alpha/K})$ the bias will likely be small.

\section{Analysis of Gene Expression Data}
\label{sec:app}

We demonstrate the utility of the shared subspace covariance
estimator for exploring differences in the covariability of gene
expression levels in young adults with different subtypes of pediatric
acute lymphoblastic leukemia (ALL) \citep{Yeoh2002}.  \edits{Quantifying biological variation across different subtypes of leukemia is important for assigning patients to risk groups, proposing appropriate treatments, and developing a deeper understanding of the mechanisms underlying these different types of cancer.  The majority of studies have focused on mean level differences between expression levels.  In particular, mean-level differences can be useful for identifying leukemia subtypes. However, differences in the covariance structure across groups can be induced by interactions between important unobserved variables.   Covariance analysis is particularly important when the effects of unobserved variables, like disease severity, disease progression or unmeasured genetic confounders, dominate mean level differences across groups.  In this analysis, we explicitly remove the mean from the data and look for differences in the covariance structure of the gene expression levels. }

\edits{The data we analyze were generated from 327 bone marrow samples analyzed on an Affymetrix oligonucleotide microarray with over 12,000 probe sets. Preliminary analysis using mean differences identified clusters corresponding to distinct leukemia subtypes: BCR-ABL, E2A-PBX1, hyperdiploid, MLL, T-ALL,
TEL-AML1.  79 patients were assigned to a seventh group for unidentified subtypes (``Others'').  We use these labels to stratify the observations into seven groups with corresponding sample sizes of $n = (15, 27, 64, 20, 43, 79, 79)$.}

Although there are over 12,000 probes on the microarray, the vast majority of gene expression
levels are missing.  Thus, we restrict our attention to the genes for
which less than half of the values are missing and use Amelia, a
software package for missing value imputation, to fill in the
remaining missing values \citep{Amelia}.  \edits{Amelia assumes the data is missing at random and that each group is normally distributed with a common covariance matrix.  Since imputation is done under the assumption of covariance homogeneity, any inferred differences between groups are unlikely to be an artifact of the imputation process.  We leave it to future work to incorporate missing data imputation into the shared subspace inference algorithm.} After removing genes with very high percentages of missing values, $p=3124$ genes remain.  Prior to analysis, we de-mean both the rows and columns of the gene expression levels in each group.

\edits{We apply the rank selection criteria discussed in Section \ref{sec:misspec} and proposed by \citet{Gavish2014} to the pooled expression data (i.e. data from all groups combined) to decide on an appropriate value for the shared subspace. This procedure yields $s=45$ dimensions\footnote{Note that for some groups, $n_k < 45$, in which case we infer the rank $r=\text{min}(n_k, s)$ $s \times s$ matrix $\Psi_k$.}. We run Algorithm \ref{alg:em} to estimate the shared subspace, and then use Bayesian inference (Algorithm \ref{alg:gibbs}) to identify differences between groups on the inferred subspace. Together, the run time for the full empirical Bayes procedure (both algorithms) took less than 10 minutes on a 2017 Macbook Pro.}

\edits{Using the goodness of fit metric, we find that a 45-dimensional shared subspace dimension that explains over $90\%$ of the estimated
variation in the top $s$ eigenvectors of $\Sigma_k$, suggesting that the rank selection procedure worked reasonably well (Figure \ref{fig:leukemiaRatio}, Appendix B). To further validate the utility of shared subspace modeling, we look at how informative the projected data covariance matrices are for predicting group membership. For an observation $Y_i$, we compute the probability, assuming uniform prior distribution over group membership, that $Y_i$ came from group $k$ as $P(Y_i \text{ from group } k ) =\frac{|\Psi_k|^{-1/2}\etr(-1/2(Y_i\hat V)^T\Psi_k^{-1}Y_i\hat V)}{\sum_j\left( |\Psi_j|^{-1/2}\etr(-1/2(Y_i\hat V)^T\Psi_j^{-1}Y_i\hat V)\right)}$.  We correctly identified the leukemia type in all samples, which provides further confirmation that this subspace provides enough predictive power to easily distinguish groups.}

\edits{In addition, we quantified differences amongst the projected data covariances using the Frobenius norm, $||\Psi_k - \Psi_j||_F$ for all pairs of the seven groups.  We use these distances to compute a hierarchical clustering dendrogram of the groups (Figure \ref{fig:dendro}, Appendix B).   The hierarchical clustering reveals that BCR-ABL, E2A-PBX1, TEL-AML1 and hyperdiploid, which correspond to B lineage leukemias, cluster together.  T-ALL, the T lineage leukemia, and MLL, the mixed lineage leukemia, appear the most different \citep{Dang2012}.  To further verify that the inferred subspace relates to relevant biological processes, we conducted gene set enrichment analysis using the observed magnitudes of the loadings for the genes on the 45 basis vectors \citep{subramanian2005gene} and using gene sets defined by the Gene Ontology Consortium \citep{GO}.  Gene set analysis on the magnitudes of gene loadings identified dozens of pathways (FDR $< 0.01$, \citep{storey2003positive}). Nearly every identified pathway relates to the immune response or cell growth (Figure \ref{fig:go}, Appendix B), for example B and T cell proliferation (GO:0042100, GO:0042102), immunoglobin receptor binding (GO:0034987) and cellular response to cytokine stimulus (GO:0071345) to name only a few.   Together, all of these results suggest that in this application there is indeed significant differences in the covariability between genes for each the of groups, with biologically plausible underpinnings.  Consequently, there is value in exploring what underlies those differences.}

\begin{figure}[t]
    \centering
    \subfigure[Subspace for hyperdiploid subtype]{
        \label{fig:leukemiaPosterior_a}
        \includegraphics[width=0.45\textwidth]{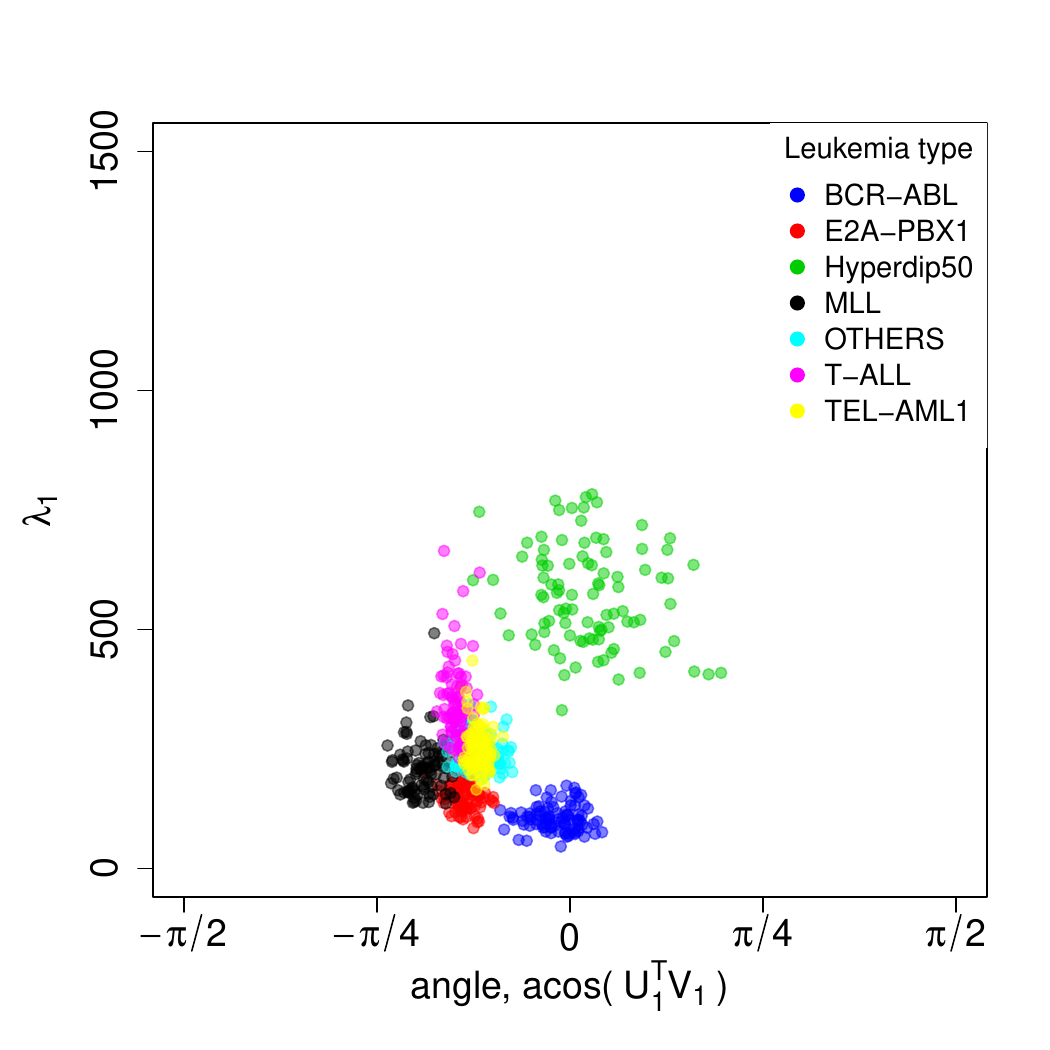}}
    ~ 
    \subfigure[T-ALL vs MLL subspace]{
      \label{fig:leukemiaPosterior_b}

    \includegraphics[width=0.45\textwidth]{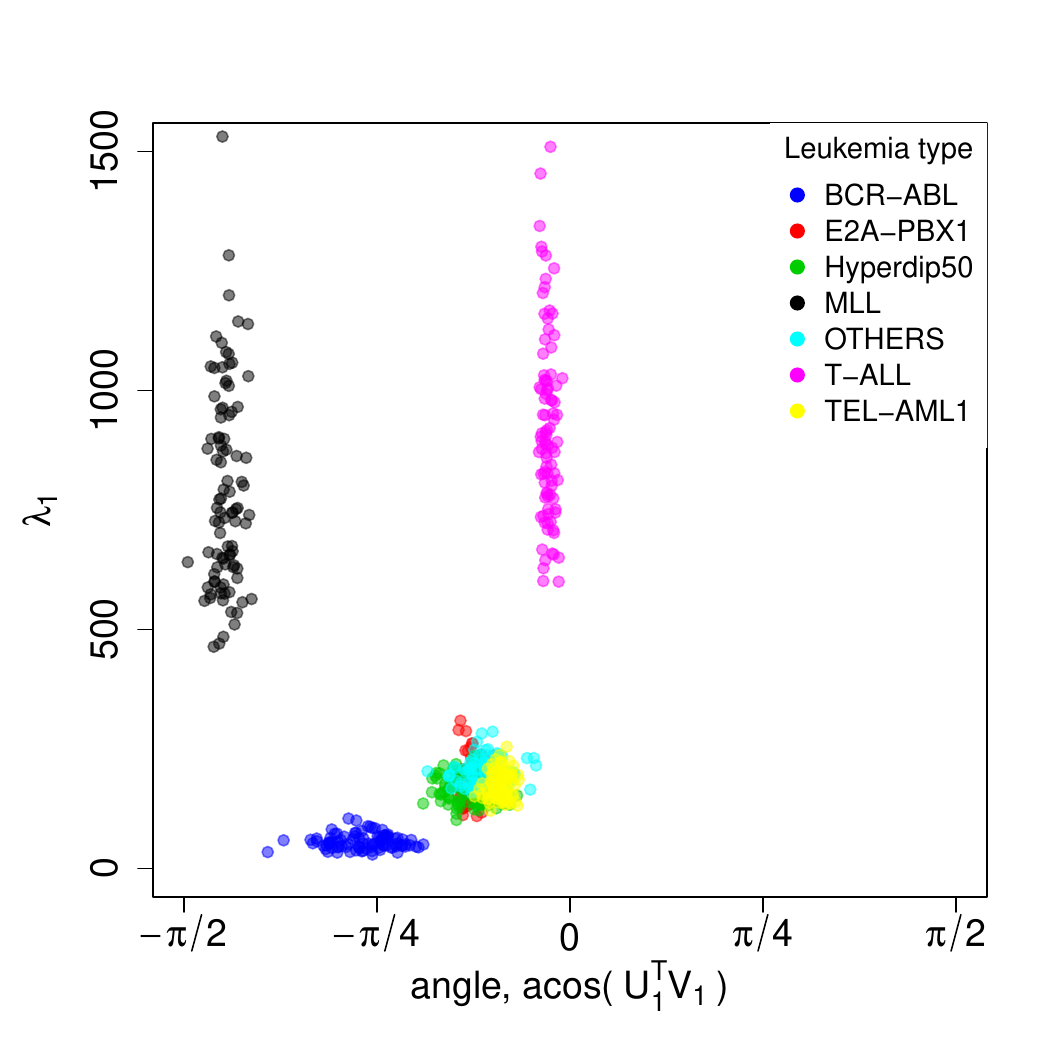}}
      \caption{Posterior samples for the
        first eigenvalue and orientation of the first eigenvector on the a dimensional subspace. a) The two dimensional subspace was chosen to approximately span the first two eigenvectors for the hyperdiploid group. The orientation of first eigenvector is similar for all groups, but the variance significantly larger for the hyperdiploid subgroup.  b) The two dimensional subspace was chosen to maximize the difference between the T-ALL and MLL groups.  Along the first dimension of this subspace, there is large variability in the T-ALL group that is not matched in other groups, whereas the second dimension there is large variability in the MLL group that is not matched in the other groups.}
\label{fig:leukemiaPosterior}
\end{figure}

We next demonstrate how we can explore significant \textit{a posteriori} differences between the groups which might lead to scientifically meaningful insights.  In order to visualize differences in the posterior distributions of the $45 \times 45$ dimensional matrices $\Psi_k$, we examine the  distribution of eigenvalues and eigenvectors between the groups on a variety of two-dimensional subspaces of the shared space. We propose two different methods for identifying potentially interesting sub-subspaces to visualize.  First, we summarize variation on a two dimensional subspace whose axes are approximately aligned to the first two eigenvectors of $\hat \Sigma_k$, for a specific group $k$. This subspace corresponds to the subspace of maximal variability within group $k$.  For example, in Figure \ref{fig:leukemiaPosterior_a} we plot posterior summaries about the principal eigenvector and eigenvalues for each group on a two dimensional space spanned by the first two eigenvectors of the inferred covariance matrix for the hyperdiploid group.  The $x$-axis corresponds to the orientation of the first eigenvector and the y-axis corresponds the magnitude of the first eigenvalue.  In this subspace, we can see that the first eigenvector for most groups appear to have  similar orientations, but that the hyperdiploid group has significantly larger variance along the first principal component direction than all other groups  (with the exception of perhaps T-ALL, for which the posterior samples overlap).  The first eigenvector for the BCR-ABL subgroup appears to be the least variable on this subspace.  

As an alternative approach to summarizing the posterior distribution, we examine the posterior eigen-summaries on a two dimensional subspace which is chosen to maximize the difference between any two chosen groups.  To achieve this, we look at spaces in which the axes correspond to the first two eigenvectors of $\hat \Sigma_k - \hat \Sigma_j$ for any $k \neq j$.  As an example, in Figure \ref{fig:leukemiaPosterior_b} we plot posterior summaries corresponding to the subspace for which the difference between the T-ALL and MLL subgroups is large.  On this subspace, the groups cluster into four distinct subgroups which appear significantly different \textit{a posteriori}: the T-ALL subtype, the MLL subtype, the BCR subtype and the all other groups.  Roughly, along the first dimension, there is large variability in the T-ALL group that is not matched in other groups, whereas the second dimension there is large variability in the MLL group that is not matched in the other groups.  

\begin{figure}[!ht]
    \centering
    \includegraphics[width=0.45\textwidth]{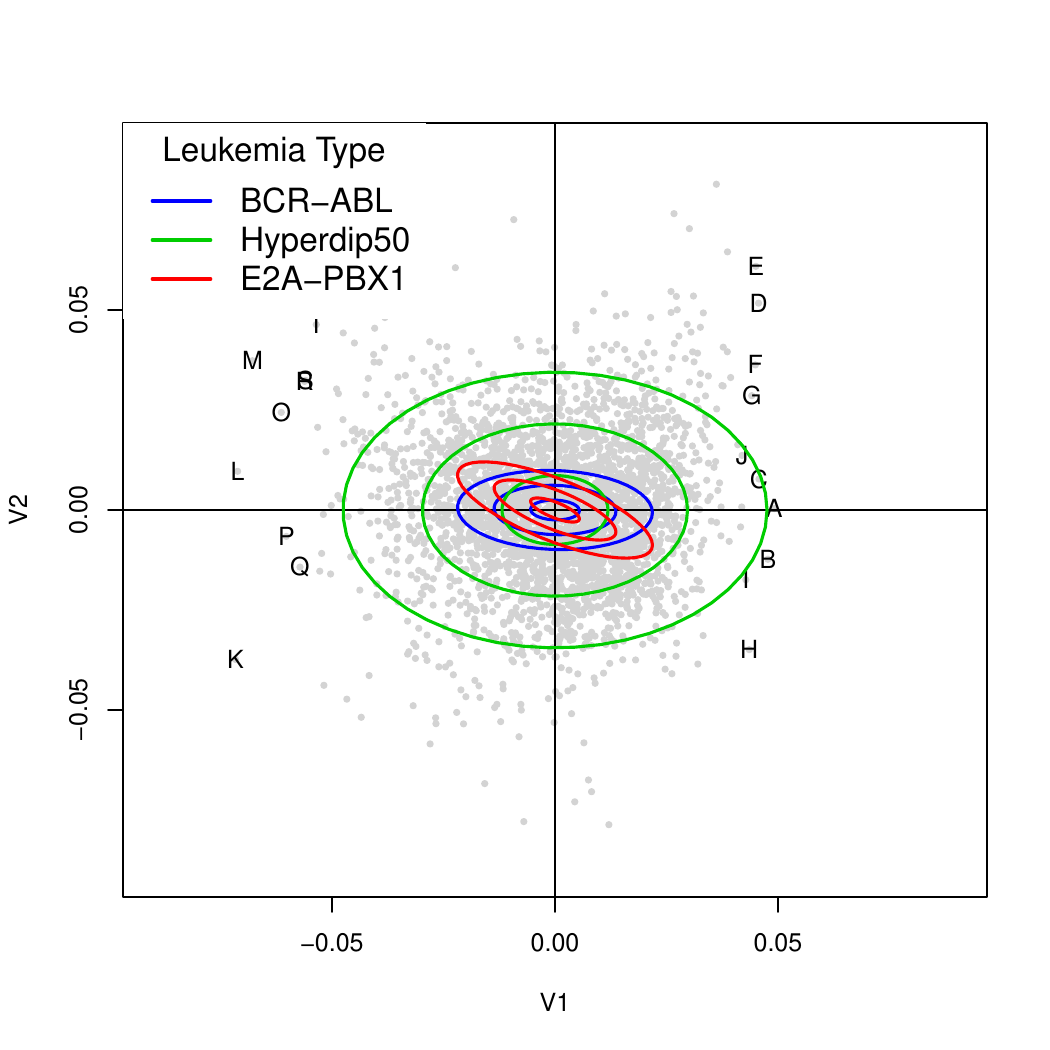}
    \qquad
\raisebox{1.2\height}{
\scriptsize
\input{Figs/genes-3-6.txt}
}
\caption{Left) Variant of a biplot for the hyperdiploid subspace.  We include contours for three leukemia
  subtypes and the loadings for each gene on the first two columns of
  $\hat{V}$.  We plot contours for three leukemia
  subtypes and the loadings for genes with the most postive (A-J) and most negative (K-T) values on the first principal axis.  The loadings for all of the genes are displayed in light gray.  There is significant correlated variability amongst genes A-T in the TEL and hyperdiploid subgroups, and a factor of two less variability amongst these genes in the E2A subgroup. Right) List of the gene's with the largest
  loadings along the first axis.}
\label{fig:leukemiaBiplot1}
\end{figure}
\begin{figure}[!h]
    \centering
    \includegraphics[width=0.45\textwidth]{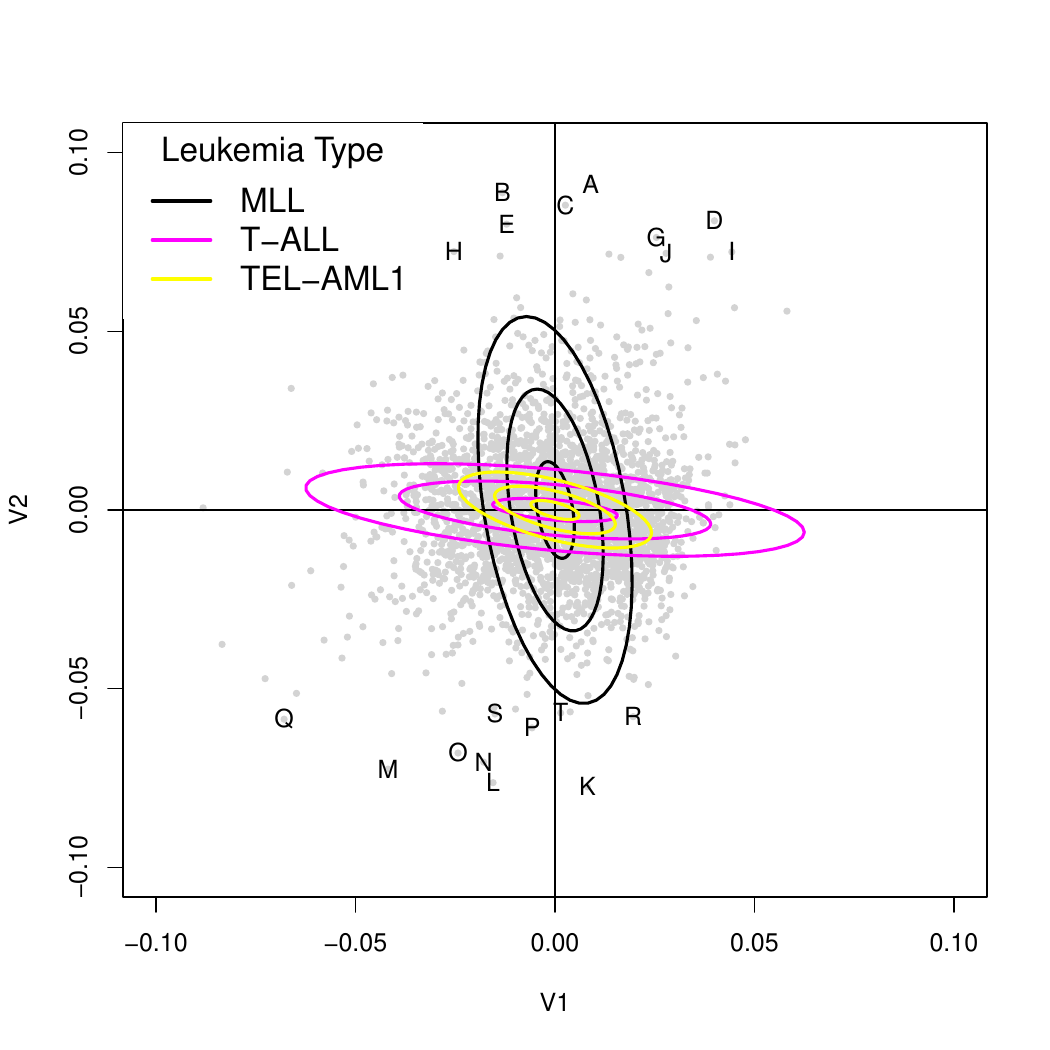}
    \qquad
\raisebox{1.2\height}{
\scriptsize
\input{Figs/genes-4-6.txt}
}
\caption{Left) Variant of a biplot for the MLL vs TEL-AML1 subspace.  We plot contours for three leukemia
  subtypes and the loadings for genes with the most positive (A-J) and most negative (K-T) values on the second axis.  The loadings for all of the genes are displayed in light gray.  There is significant correlated variability among genes with large loadings (e.g. letters A through T) in the MLL subgroup, and a significantly less variability in the TEL-AML1 and T-ALL groups.  Although the TEL and T-ALL groups have similar variance in the ``V2'' direction, T-ALL has significantly more variance in the ``V1'' direction.  Right) List of the gene's with the largest loadings along the V2 axis.}
\label{fig:leukemiaBiplot7}
\end{figure}

Scientific insights underlying the significant differences that were identified in Figure \ref{fig:leukemiaPosterior} can be understood in the biplots in Figures \ref{fig:leukemiaBiplot1} and \ref{fig:leukemiaBiplot7}.  In each figure, we plot the contours of the two dimensional covariance matrices for a few leukemia subtypes. The 20 genes with the largest loadings for one of the component directions are indicated with letters and the remaining loadings plotted with light grey dots.  The gene names for the genes with the largest loadings are listed in the corresponding table.  In both biplots, the identified genes have known connections to cancer, leukemia, and the immune system. 

For example, for the subspace of maximal variability in the hyperdiploid group, gene set analysis identified two gene sets with large magnitude loadings on the first principal component: a small group of proteins corresponding to the MHC class II protein complex (GO:0006955) as well as a larger group of genes corresponding to genes generally involved in immune response (GO:0006955).  MHC class II proteins are known to play an essential role in the adaptive immune system \citep{reith2005regulation} and are correlated with leukemia patient outcomes \citep{rimsza2004loss}.  Our analysis indicates these proteins have especially variable levels in the hyperdiploid subtype relative to the other leukemia subtypes.

\edits{For the subspace chosen to maximize the difference between T-ALL and MLL groups, gene set analysis associated with large loadings in the second dimension (associated with high variance in the MLL subgroup) included ``regulation of myeloid cell differentiation'' (GO:0045637), ``positive regulation of B cell receptor signaling pathway'' (GO:0098609) and ``immunoglobulin V(D)J recombination'' (GO:0033152). Most of the individual genes with large loadings are known in the leukemia literature including WASF1 (``F'') which plays an important role in apoptosis \citep{kang2010wave1},  LEF1 (``D'')  which is linked to the pathogenesis of leukemia \citep{gutierrez2010inactivation} and LMO2 (``M'') which was shown to initiate leukemia in mice \citep{mccormack2010lmo2}, to name only a few.  In contrast to the MLL group, these genes in the T-ALL and TEL-AML1 subgroups have relatively little variability.}


These insights would be overlooked in more conventional mean-based analyses, particularly when mean-level differences are small relative to the residual variance.  Further, we have shown how the shared subspace reveals sets of interpretable genes that are most important for describing difference between leukemia subtypes; these discoveries would less evident with alternative covariance estimation methods which do not explicitly include the assumption about differences manifesting on a common low dimensional subspace.  All told, these results highlight the value of shared subspace covariance matrix inference for both predicting leukemia subtypes as well as for exploring scientifically meaningful differences between the groups.

\section{Discussion}

In this paper, we proposed a class of models for estimating and
comparing differences in covariance matrices across multiple groups on
a common low dimensional subspace.  We described an empirical Bayes
algorithm for estimating this common subspace and a Gibbs sampler for
inferring the projected covariance matrices and their associated
uncertainty.  Estimates of both the shared subspace and the projected
covariance matrices can both be useful summaries of the data.  For
example, with the leukemia data, the shared subspace highlights the
full set of genes that are correlated across groups.
Differences between group covariance matrices can be understood in
terms of differences in these sets of correlated molecules.  In this
analysis, we demonstrated how we can use these notions to visualize and contrast the posterior distributions of covariance matrices projected onto a particular
subspace and interpret these differences biologically.

In simulation, we showed that the shared subspace model can still be a
reasonable choice for modeling multi-group covariance matrices even
when the groups may be largely dissimilar.  When there is little
similarity between groups, the shared subspace model can still be appropriate as
long as the dimension of the shared subspace is large enough.
However, selecting the rank of the shared subspace remains a practical
challenge.  Although we propose a useful heuristic for choosing the
dimension of the shared subspace based on the rank selection
estimators of \citet{Gavish2014}, a more principled approach is
warranted.  Improved rank estimators would further improve the
performance of the adaptive shared subspace estimator discussed in
Section \ref{sec:simulation}.

It is also a challenging problem to estimate the ``best'' subspace
once the rank of the space is specified.  We used maximum marginal
likelihood to estimate $VV^T$ and then used MCMC to
infer $\Psi_k$.  By focusing on group differences for $\Psi_k$ on a
\emph{fixed} subspace, it is much simpler to interpret similarities
and differences.  Nevertheless, full uncertainty quantification for
$VV^T$ can be desirable.  We found MCMC inference for $VV^T$ to be
challenging for the problems considered in this paper and leave it for
future work to develop an efficient fully Bayesian approach for
estimating the joint posterior of $VV^T$ and $\Psi_k$.  Recently
developed Markov chain Monte Carlo algorithms, like Riemannian
manifold Hamilton Monte Carlo, which can exploit the geometry of the
Grassmannian manifold, may be useful here \citep{Byrne2013,
  Girolami2011}.  It may also be possible, though computationally
intensive, to jointly estimate $s$ and $VV^T$ using for instance,
a reversible-jump MCMC algorithm.

Fundamentally, our approach is quite general and can be integrated
with existing approaches for multi-group covariance
estimation.  In particular, we can incorporate additional shrinkage on
the projected covariance matrices $\Psi_k$.  As in \citet{Hoff2009} we
can employ non-uniform Bingham prior distributions for the
eigenvectors of $\Psi_k$ or we can model $\Psi_k$ as a function of continuous
covariates as in \citet{Yin2010} and \citet{Hoff2011}.  Alternatively, we can
summarize the estimated covariance matrices by thresholding entries of
the precision matrix, $\Psi_k^{-1}$ to visualize differences between
groups using a graphical model \citep{Meinshausen2006}.  \edits{We can also incorporate sparsity to the estimated eigenvectors of the shared subspace to add in interpretation \citep[e.g]{rovckova2016fast}.  Finally, we can consider variants in which some eigenvectors are assumed to be identical across groups, whereas others are allowed to vary on the shared subspace.  This can further improve estimation efficiency, particularly when the common eigenvectors are associated with the largest eigenvalues and differences appear in lower variance components \citep{cook2008covariance}.  Such an approach would further aid in identifying the relevant sub-subspace of variability that describes prominent differences between groups .} The specifics
of the problem at hand should dictate which extensions are
appropriate, but the shared subspace assumption can be useful in a
wide range of analyses, especially when the number of features is very
large.  A repository for the replication code is available on GitHub \citep{FranksGit}.



\newpage

\appendix

\section{Additional Misspecification Results}

\begin{figure}[t]
    \centering
    \subfigure[]{
      \label{fig:a}
      \includegraphics[width=0.3\textwidth]{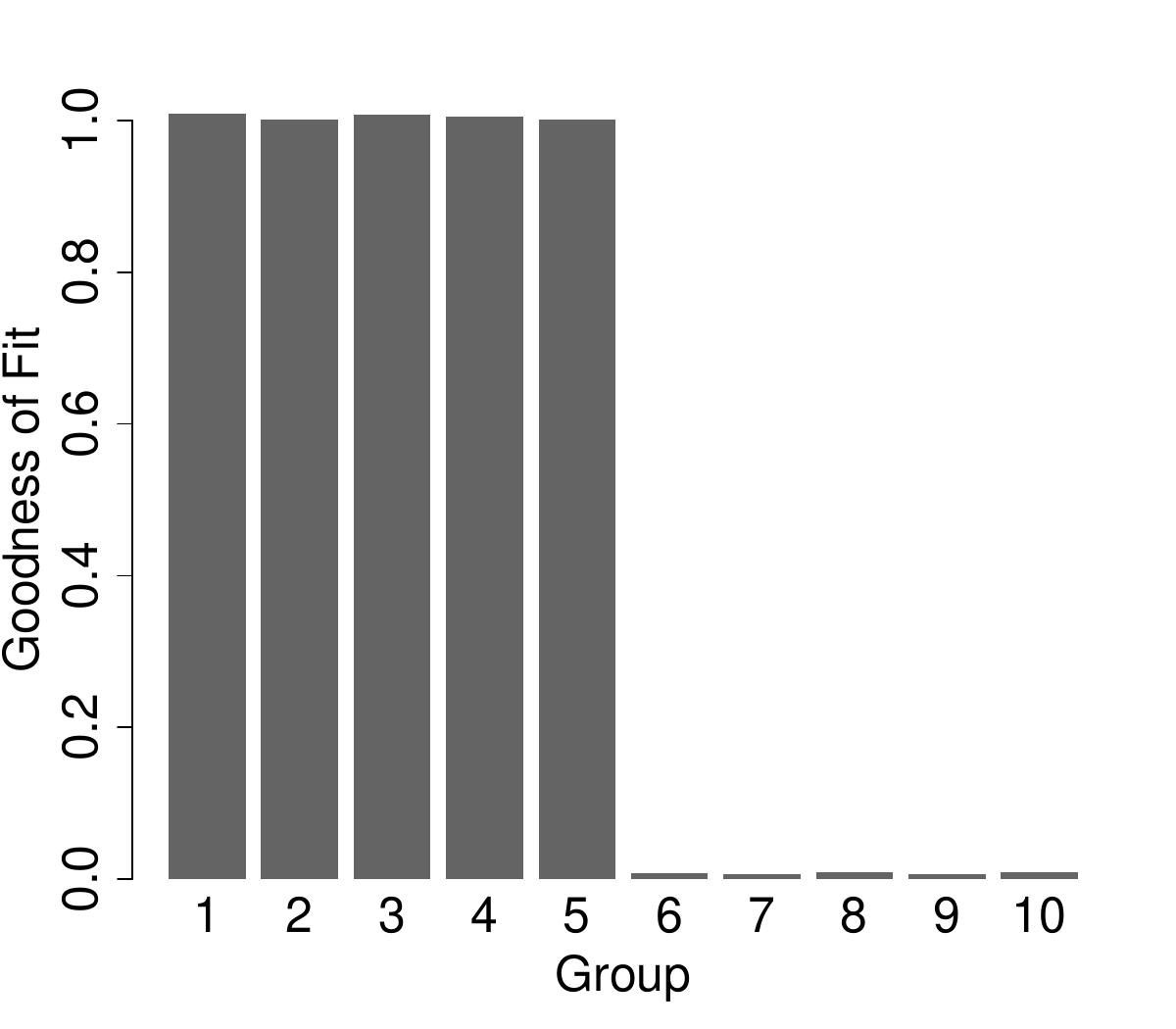}}
    \subfigure[]{
      \label{fig:b}
      \includegraphics[width=0.3\textwidth]{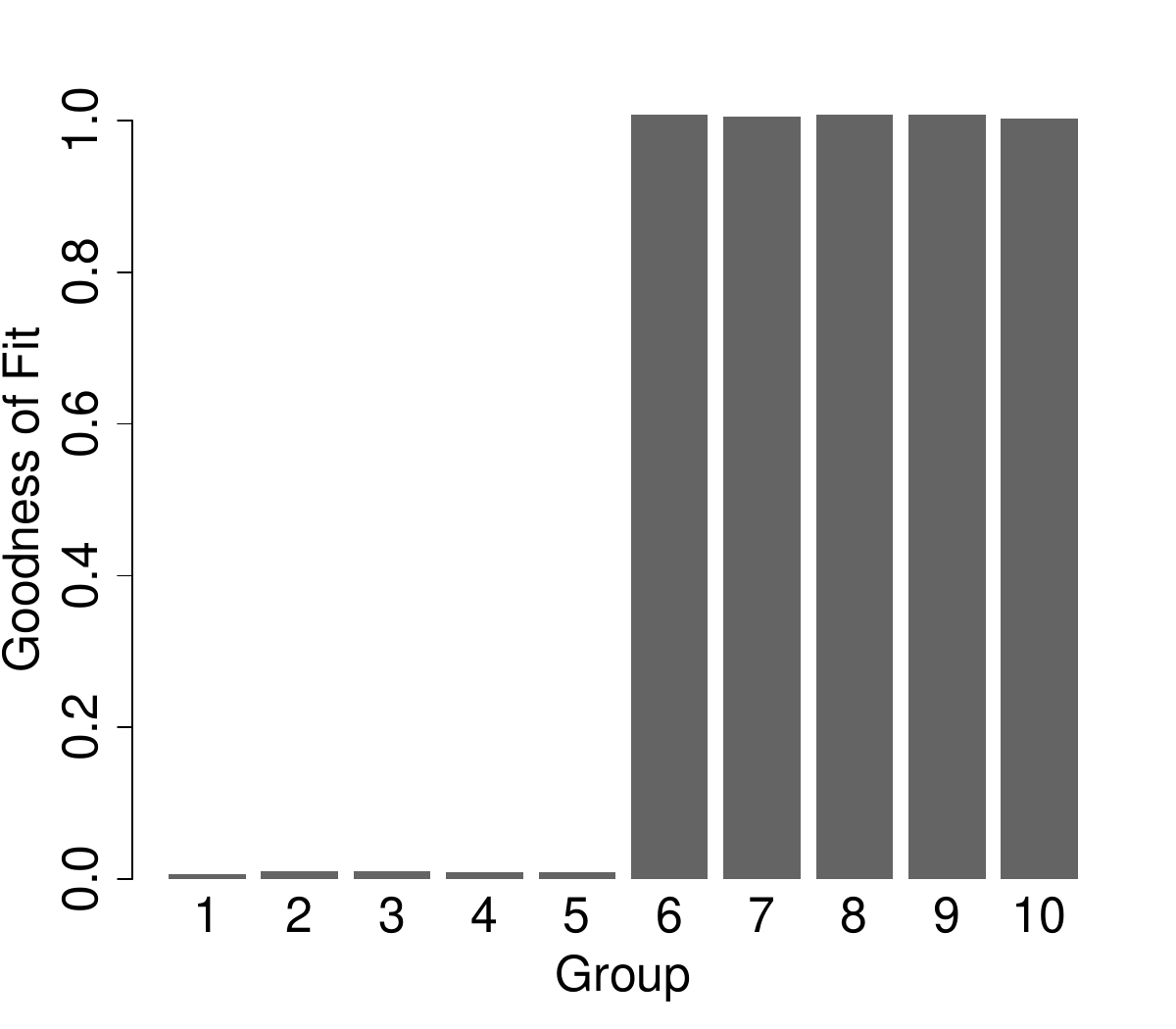}}
    \subfigure[]{
        \label{fig:c}
        \includegraphics[width=0.3\textwidth]{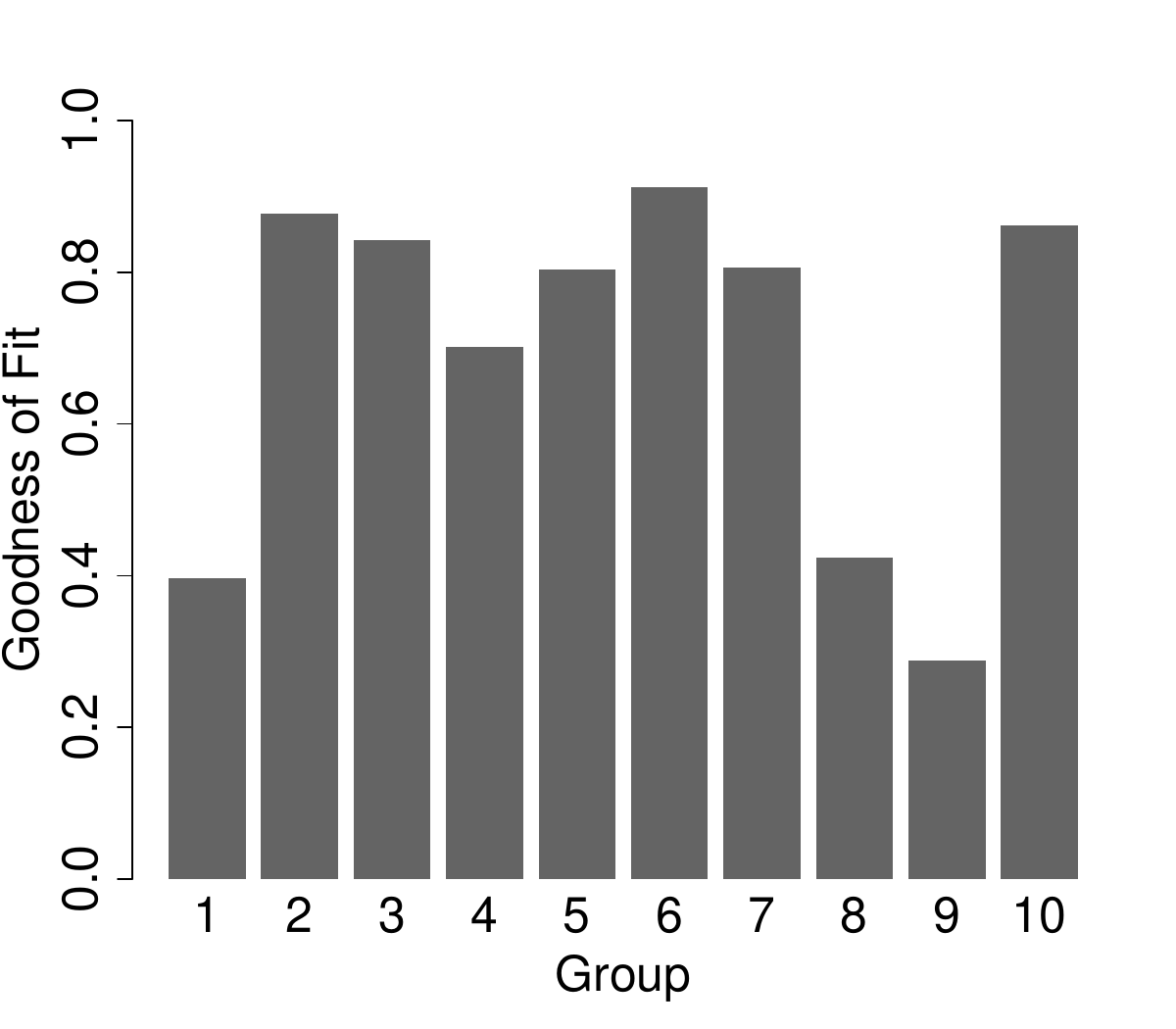}}
      \caption{Goodness of fit for 10 groups using a two dimensional shared subspace model.  In truth, the eigenvectors of the first five groups share a 2 dimensional subspace and the eigenvectors of the last five groups share a different 5 dimensional subspace.  Empirically, by initializing the shared subspace uniformly at random, we found that there were three local modes. a) This mode corresponds to the shared subspace of the first five groups.  b) This mode corresponds to the shared subspace of the second five groups.  c) The third mode corresponds to a ``shared subspace'' across all groups. This is the mode discovered when using the eigen-based initialization strategy suggested in Section \ref{sec:misspec}. Note that in truth variation in all ten groups could be captured using a 4 dimensional shared subspace.}
\label{fig:five_share}
\end{figure}

Following the simulation set up of \ref{sec:misspec} we generate data from 10 groups with $(\lambda_1, \lambda_2) = (250, 25)$, $p =200$ and $\sigma_k^2=1$.  In this section, we consider two model misspecification simulations.  First, we consider data in which the first two eigenvectors for the first five groups share a two-dimensional subspace, and the eigenvectors for the last five groups share a different two-dimensional subspace.  We then fit all ten groups assuming a two-dimensional shared subspace model.  In Figure \ref{fig:five_share} we plot the goodness of fit metric for all ten groups for subspaces identified in different local modes of the likelihood.  Specifically, we empirically identified three local modes: one mode identifies the shared subspace for the first give groups, the other mode corresponds to the shared subspace for the second five groups, and the third mode corresponds to subspace shares some commonalities across all 10 groups.  This last mode is the one discovered by the eigen-based initialization strategy proposed in Section \ref{sec:misspec}.  

\begin{figure}[h]
    \centering
    \subfigure[]{
      \label{fig:d}
      \includegraphics[width=0.3\textwidth]{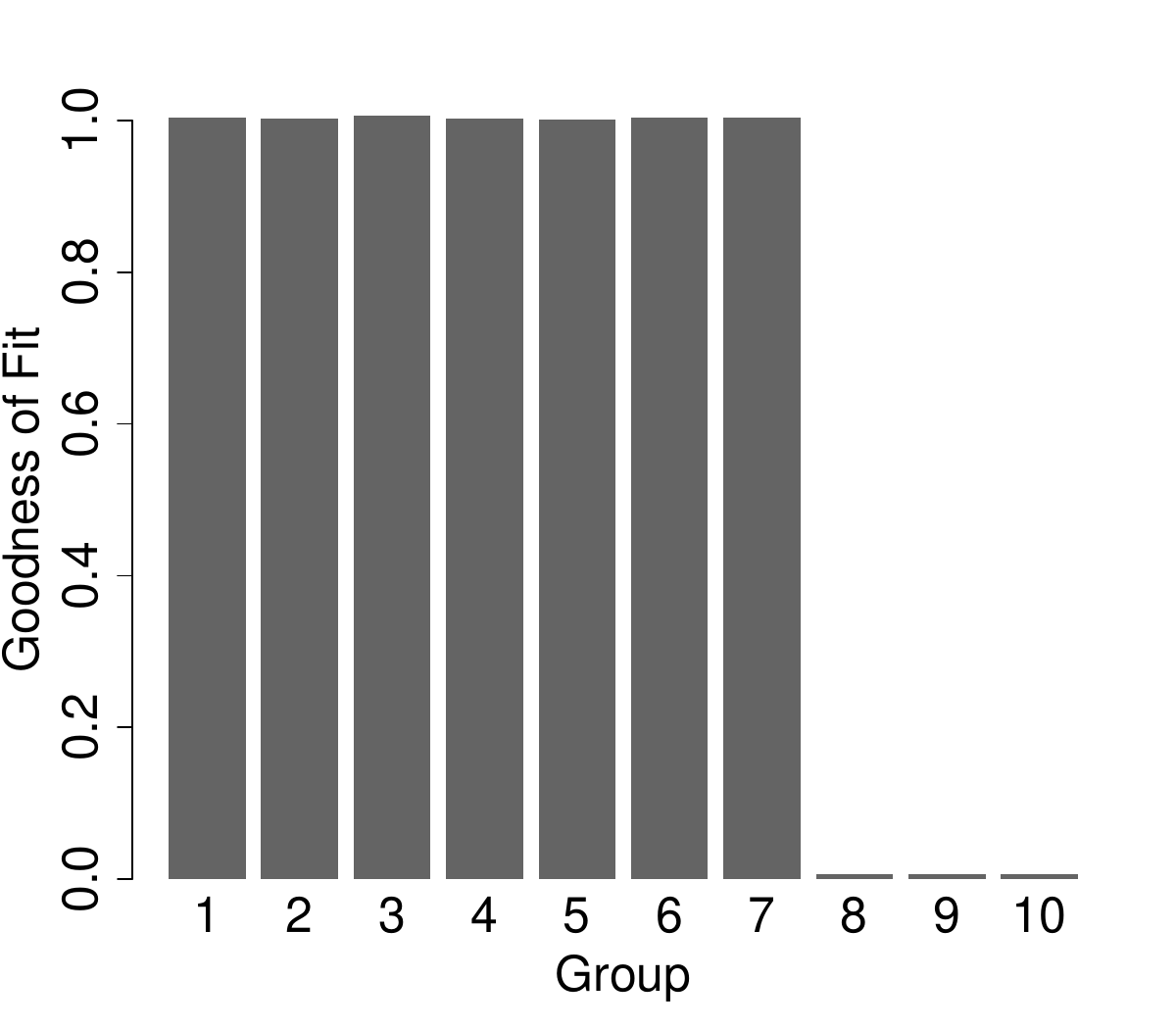}}
    \subfigure[]{
      \label{fig:e}
      \includegraphics[width=0.3\textwidth]{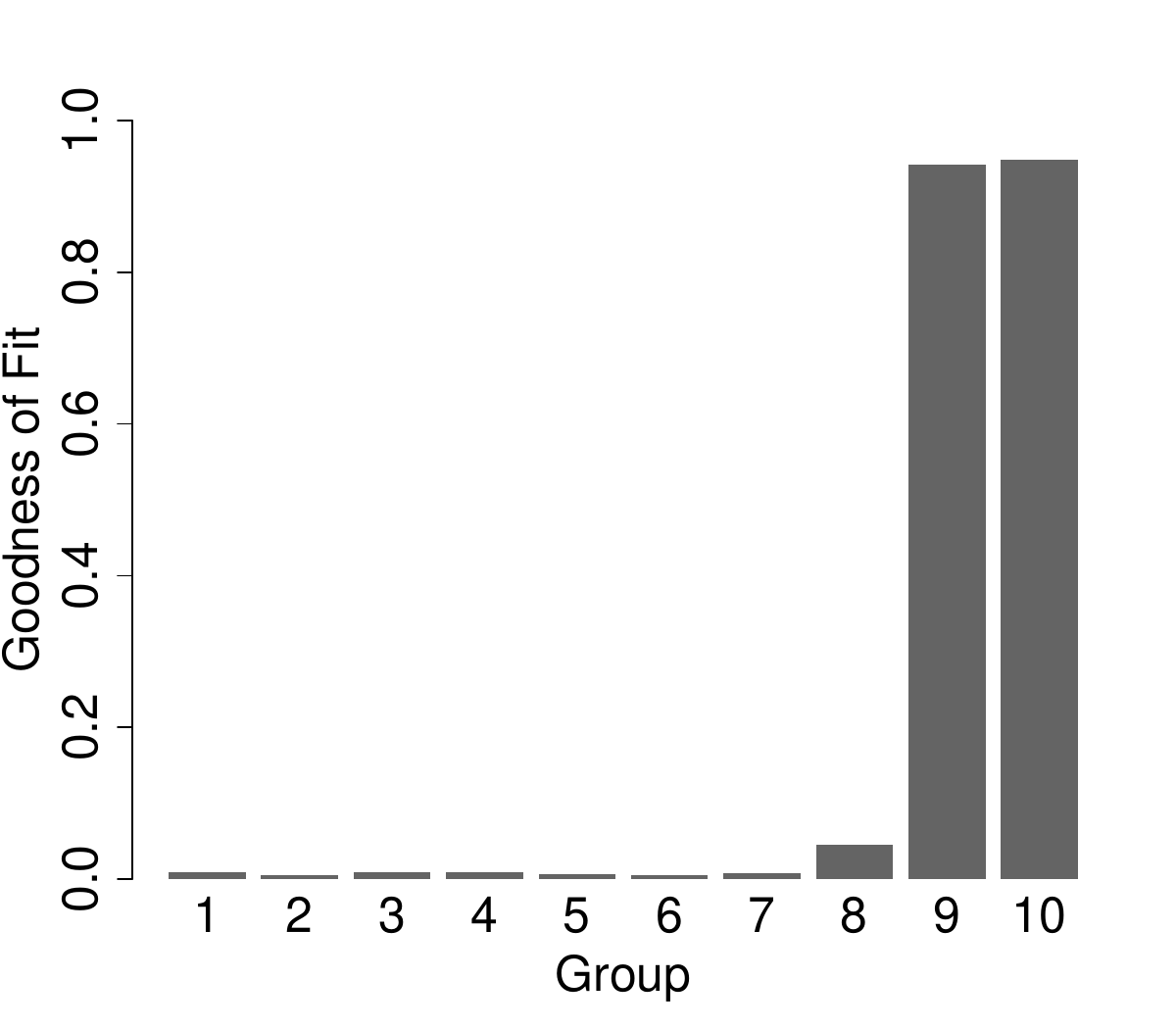}} \\
    \subfigure[]{
        \label{fig:f}
        \includegraphics[width=0.3\textwidth]{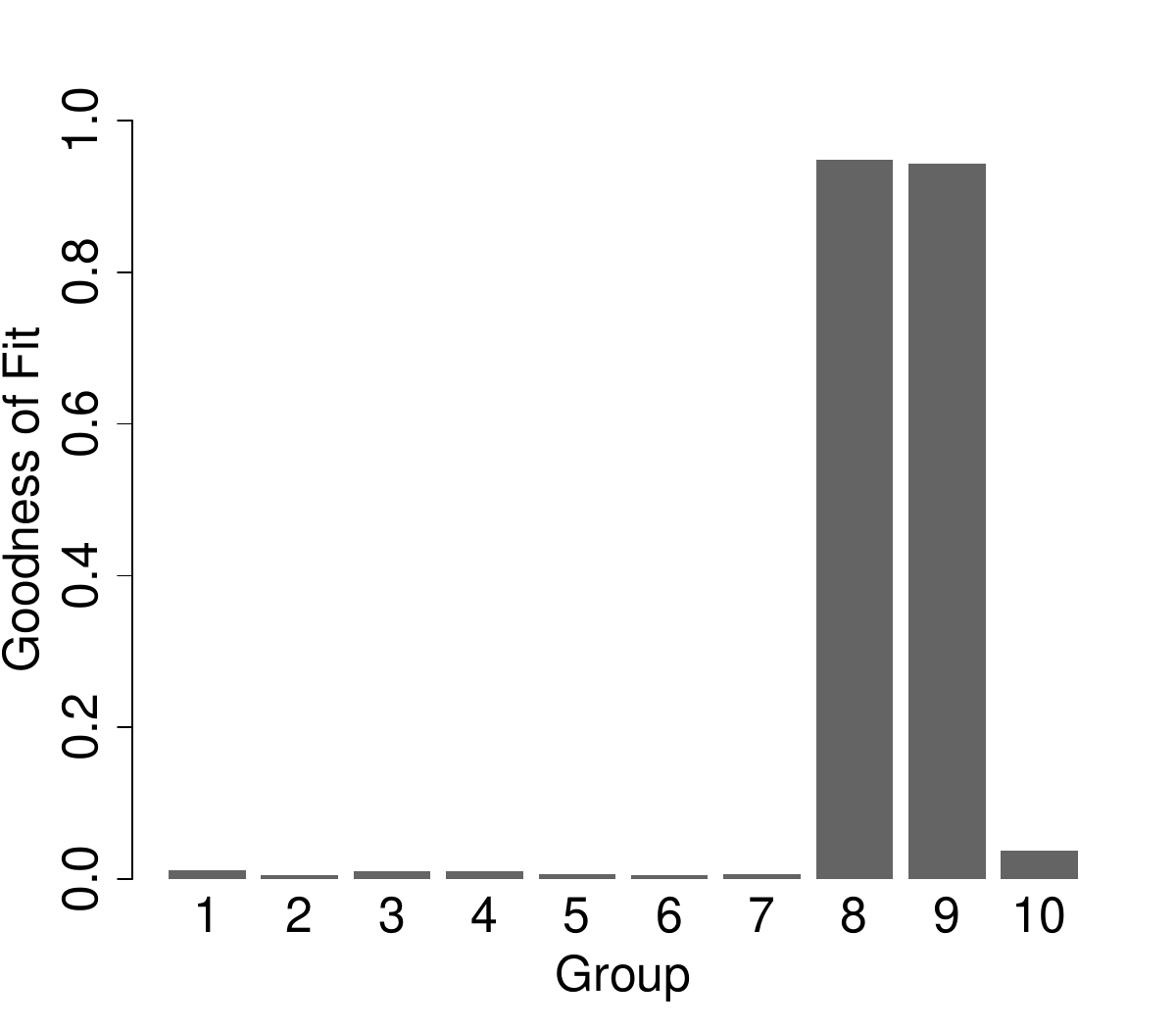}}
    \centering
    \subfigure[]{
      \label{fig:g}
       \includegraphics[width=0.3\textwidth]{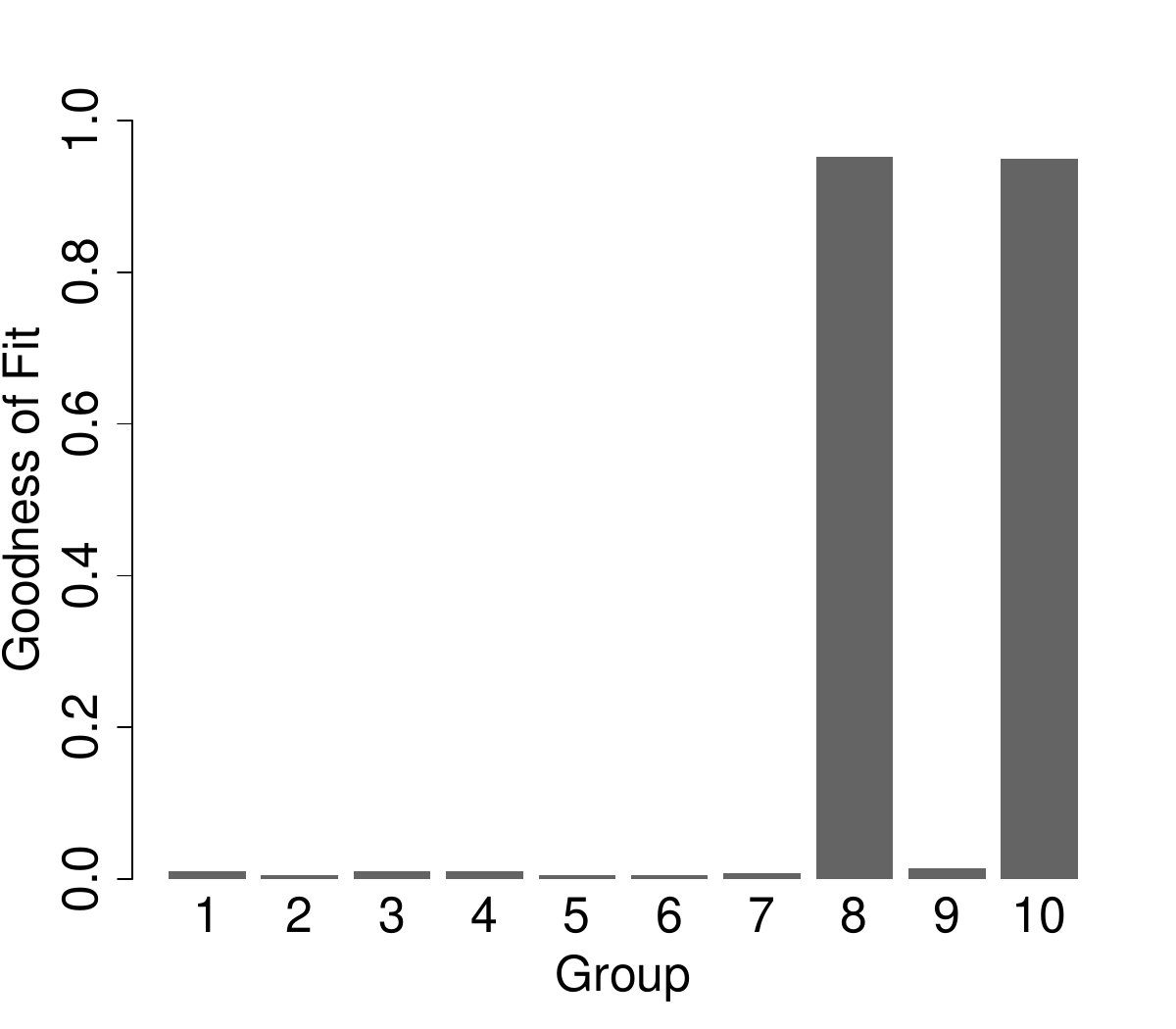}}
      \caption{Goodness of fit for 10 group shared subspace model.  The eigenvectors of the first seven groups share a 2 dimensional subspace and the eigenvectors of the last three groups were generated uniformly on the null space.  Empirically, by initializing the shared subspace uniformaly at random, we found that there were three local modes. a) We discover the subspace shared by the first seven groups using the eigen-based initialization (Section \ref{sec:misspec}).  We also identify that the last three groups have small variance on this subspace, indicating that they do not share the subspace. b-d) Additional local (non-global) modes can be identified which in which 2 of the last three groups approximately share a two-dimensional subspace.}      
      
\label{fig:seven_share}
\end{figure}

In the second simulation we generate the first two eigenvectors for the first seven groups from a common two dimensional subspace.  The eigenvectors from the last three groups were generated uniformly at random from the $p-2$ dimensional null space of the shared subspace. In repeated simulations with $V$ initialized uniformly at random on the Stiefel manifold, the resulting we empirically discovered four modes.  In  Figure \ref{fig:seven_share} we plot goodness of fit metrics for the 10 groups at these modes.  The first mode corresponds to the shared subspace for the first 7 groups.  The other three modes identify subspaces shared by two of the last three groups.

\begin{figure}[t]

\end{figure}

\clearpage
\newpage

\section{Run time results}
\begin{figure}[h]
\centering
     \includegraphics[width=0.8\textwidth]{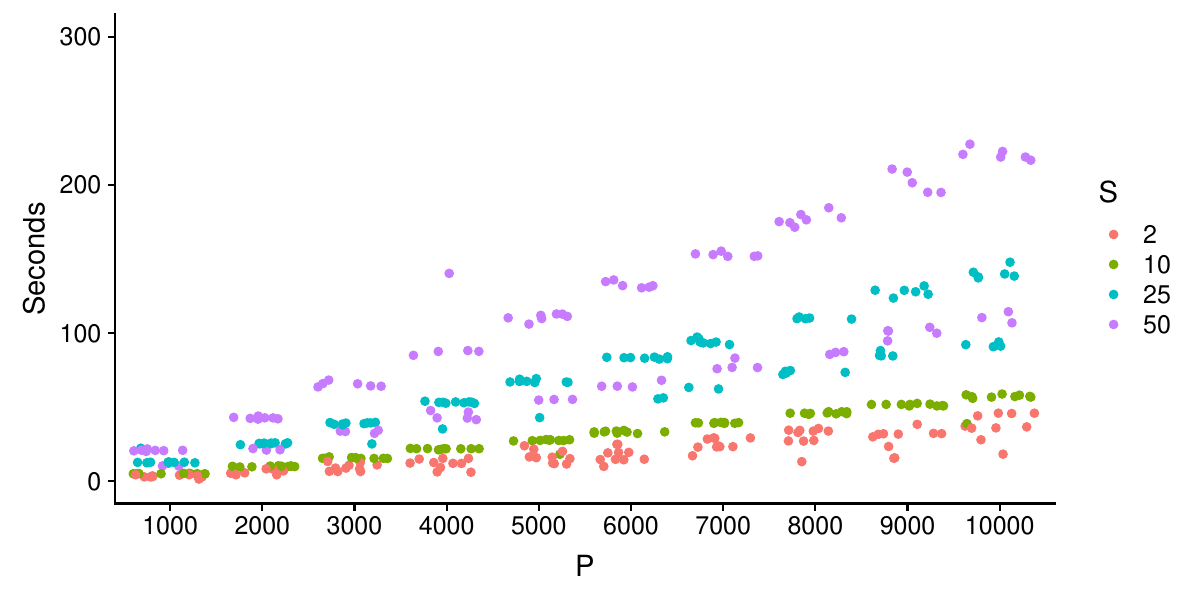}
      \caption{Run time results for subspace inference (Algorithm \ref{alg:em}) as a function of the subspace dimension, $S$, and the number of features $p$. $s = {2, 10, 25, 50}$ and $p = 1000, 2000, ... 10000$.  Points are jittered for visibility.  In this simulation we assume $K=5$ groups, $n_k=50$ observations per group, $\sigma^2_k =1$ and the eigenvalues of $\psi_k$ are samples from an Expo(1/4) (e.g. have mean 4).  For each value of $S$ and $P$ we run subspace inference 10 times and plot the resulting run times. In each simulation we initialize the optimization routine uniformly at random on $\mathcal{V}_{p, s}$ to get a conservative estimate for run times.  Using the intelligent initialization routine discussed in Section \ref{sec:misspec} typically increases time to convergence.  Convergence time is on the order of minutes, even for relatively large values of $s$ and $p$.}
\label{fig:runtimes}
\end{figure}

\section{Addition Results From Leukemia Analysis}

\begin{figure}[h]

    \centering
    \subfigure[$\gamma(Y_k: \hat{V}, \hat{\sigma_k}^2)$]{
      \label{fig:leukemiaRatio}
         \includegraphics[width=0.4\textwidth]{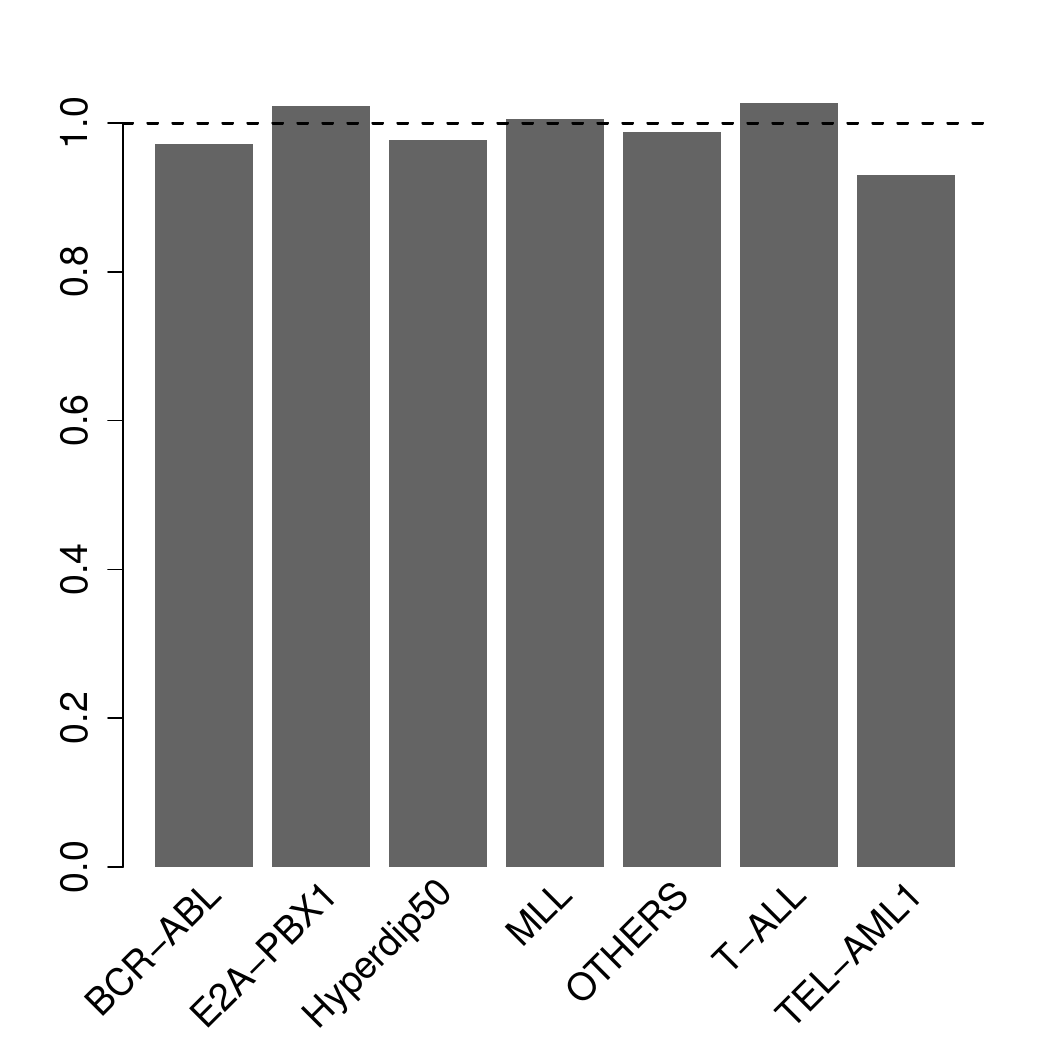}}
         \qquad
    \subfigure[Hierarchical clustering]{
      \label{fig:dendro}
         \includegraphics[width=0.4\textwidth]{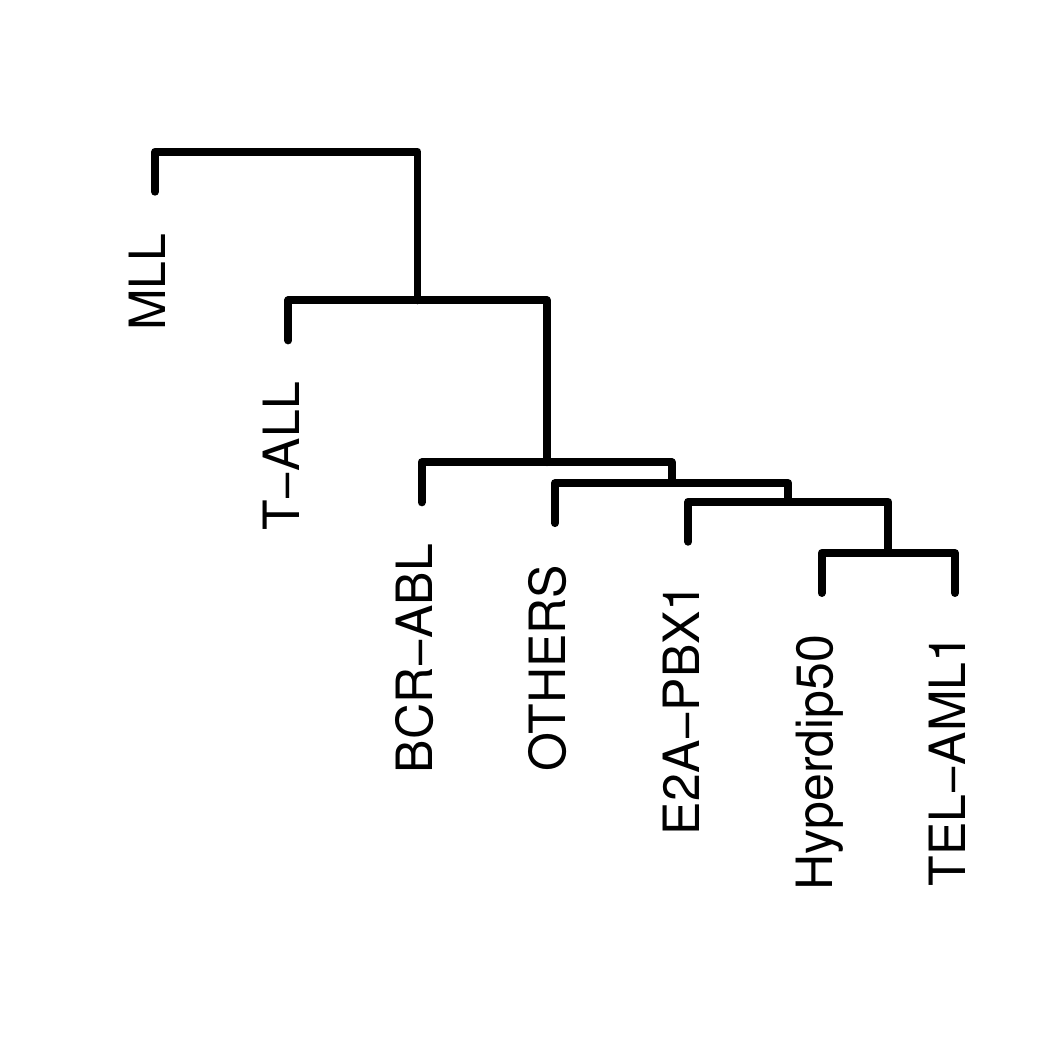}}
      \caption{\edits{a) Goodness of shared subspace fit for each of the
        seven Leukemia groups.  The inferred $s=45$ dimensional subspace explains over $90\%$ of estimated total variation in $\Sigma_k$ in each of the seven groups.  b) Complete-linkage hierarchical clustering of inferred projected data leukemia covariance matrices using Frobenius norm distance metric.  The right sub-branches, which includes BCR-ABL, E2A-PBX1, hyperdiploid and TEL-AML1, are the B lineage leukemias of the seven types.  T-All is a T lineage leukemia and MLL is a mixed lineage leukemia \citep{Dang2012}. }}
\label{fig:leukemia_checks}
\end{figure}


\begin{figure}[!ht]
\centering
\raisebox{5ex}{
\tiny
\input{Figs/go-all.txt}
}
\caption{Gene set enrichment analysis based on the magnitude of gene-loadings on the inferred 45 dimensional shared subspace (Section 6).}
\label{fig:go}
\end{figure}

\vskip 0.2in
\clearpage

\bibliographystyle{plainnat}
\bibliography{refs.bib}

\end{document}

%% file: Figs/genes-3-6.txt
\begin{tabular}{ll|ll}
  \hline
 & Positive &  & Negative \\ 
  \hline
A & HLA-DQB1 & K & BCL11A \\ 
  B & HBG1 & L & HHLA1 \\ 
  C & SASH3 & M &  \\ 
  D & HLA-DPB1 & N & CSHL1 \\ 
  E & MME & O & NF2 \\ 
  F & HLA-DQB1 & P & SKAP2 \\ 
  G & DPYSL2 & Q & TRDV2 \\ 
  H & PRPF6 & R & EIF2AK2 \\ 
  I & ADA & S &  \\ 
  J & ATP6V0E2 & T & PMF1 \\ 
   \hline
\end{tabular}

%% file: Figs/genes-4-6.txt
\begin{tabular}{ll|ll}
  \hline
 & Positive &  & Negative \\ 
  \hline
A & SELL & K & AHNAK \\ 
  B & CD24 & L & NR3C1 \\ 
  C & SH3BP5 & M & LMO2 \\ 
  D & LEF1 & N & NR3C1 \\ 
  E & CCR7 & O & GSN \\ 
  F & WASF1 & P & SERPINB1 \\ 
  G & LSP1 & Q & CSHL1 \\ 
  H & FADS3 & R & DPYSL2 \\ 
  I & LCK & S & NKG7 \\ 
  J & LCK & T & DAD1 \\ 
   \hline
\end{tabular}

%% file: Figs/go-all.txt
\begin{tabular}{rlrr}
  \hline
 & Name & Q-value & Number of Genes \\ 
  \hline
GO:0005751 & mitochondrial respiratory chain complex IV & 0.00 &   7 \\ 
  GO:0044388 & small protein activating enzyme binding & 0.01 &   7 \\ 
  GO:0022624 & proteasome accessory complex & 0.00 &  16 \\ 
  GO:0048025 & negative regulation of nuclear mRNA splicing, via spliceosom & 0.01 &  21 \\ 
  GO:0004298 & threonine-type endopeptidase activity & 0.01 &  19 \\ 
  GO:0010498 & proteasomal protein catabolic process & 0.01 &  26 \\ 
  GO:0006405 & RNA export from nucleus & 0.00 &  33 \\ 
  GO:0031124 & mRNA 3'-end processing & 0.00 &  36 \\ 
  GO:0030336 & negative regulation of cell migration & 0.01 &  22 \\ 
  GO:0038083 & peptidyl-tyrosine autophosphorylation & 0.01 &  20 \\ 
  GO:0043235 & receptor complex & 0.00 &  18 \\ 
  GO:0045766 & positive regulation of angiogenesis & 0.00 &  36 \\ 
  GO:0048661 & positive regulation of smooth muscle cell proliferation & 0.00 &  18 \\ 
  GO:0035690 & cellular response to drug & 0.00 &  27 \\ 
  GO:0060337 & type I interferon-mediated signaling pathway & 0.00 &  32 \\ 
  GO:0000786 & nucleosome & 0.00 &  27 \\ 
  GO:0004888 & transmembrane signaling receptor activity & 0.00 &  27 \\ 
  GO:0030183 & B cell differentiation & 0.00 &  30 \\ 
  GO:0030890 & positive regulation of B cell proliferation & 0.01 &  15 \\ 
  GO:0060333 & interferon-gamma-mediated signaling pathway & 0.00 &  35 \\ 
  GO:0030198 & extracellular matrix organization & 0.00 &  23 \\ 
  GO:0002053 & positive regulation of mesenchymal cell proliferation & 0.01 &   9 \\ 
  GO:0071345 & cellular response to cytokine stimulus & 0.01 &  19 \\ 
  GO:0007159 & leukocyte cell-cell adhesion & 0.00 &  16 \\ 
  GO:0034113 & heterotypic cell-cell adhesion & 0.00 &  10 \\ 
  GO:0042102 & positive regulation of T cell proliferation & 0.00 &  19 \\ 
  GO:0042605 & peptide antigen binding & 0.00 &  15 \\ 
  GO:0030658 & transport vesicle membrane & 0.00 &  13 \\ 
  GO:0071556 & integral to lumenal side of endoplasmic reticulum membrane & 0.01 &  17 \\ 
  GO:0042613 & MHC class II protein complex & 0.00 &  10 \\ 
  GO:0004896 & cytokine receptor activity & 0.00 &   9 \\ 
  GO:0005001 & transmembrane receptor protein tyrosine phosphatase activity & 0.01 &   7 \\ 
  GO:0030669 & clathrin-coated endocytic vesicle membrane & 0.00 &  10 \\ 
  GO:0042100 & B cell proliferation & 0.00 &  14 \\ 
  GO:0042742 & defense response to bacterium & 0.00 &  35 \\ 
  GO:0031668 & cellular response to extracellular stimulus & 0.00 &  13 \\ 
  GO:0001916 & positive regulation of T cell mediated cytotoxicity & 0.01 &  10 \\ 
  GO:0019731 & antibacterial humoral response & 0.00 &  19 \\ 
  GO:0001915 & negative regulation of T cell mediated cytotoxicity & 0.00 &   6 \\ 
  GO:0072562 & blood microparticle & 0.00 &  32 \\ 
  GO:0035456 & response to interferon-beta & 0.01 &   7 \\ 
  GO:0050829 & defense response to Gram-negative bacterium & 0.00 &  13 \\ 
  GO:0003823 & antigen binding & 0.00 &  24 \\ 
  GO:0071757 & hexameric IgM immunoglobulin complex & 0.00 &   6 \\ 
  GO:0006911 & phagocytosis, engulfment & 0.00 &  23 \\ 
  GO:0042834 & peptidoglycan binding & 0.00 &   7 \\ 
  GO:0071756 & pentameric IgM immunoglobulin complex & 0.00 &   7 \\ 
  GO:0006958 & complement activation, classical pathway & 0.00 &  19 \\ 
  GO:0006910 & phagocytosis, recognition & 0.00 &  16 \\ 
  GO:0042571 & immunoglobulin complex, circulating & 0.00 &  16 \\ 
  GO:0050871 & positive regulation of B cell activation & 0.00 &  16 \\ 
  GO:0003094 & glomerular filtration & 0.00 &   7 \\ 
  GO:0034987 & immunoglobulin receptor binding & 0.00 &  17 \\ 
  GO:0001895 & retina homeostasis & 0.00 &  11 \\ 
   \hline
\end{tabular}